\documentclass[pra,twocolumn,showpacs,floatfix,footinbib]{revtex4}

\usepackage[final]{graphicx}
\usepackage{amsfonts,amssymb,amsmath}

\newcommand{\average}[1]{\left\langle #1 \right\rangle}

\begin{document}

\title{Dynamical properties of nanolasers based on few discrete emitters}

\author{Anders Moelbjerg}\email{amolu@fotonik.dtu.dk}
\author{Per Kaer}
\author{Michael Lorke}
\author{Bjarne Tromborg}
\author{Jesper M{\o}rk}
\affiliation{DTU Fotonik, Department of Photonics Engineering, Technical University of Denmark, Building 343, 2800 Kgs. Lyngby, Denmark}

\date{\today}

\begin{abstract}
We investigate the dynamical properties of nanolasers comprising a few two-level emitters coupled to an optical cavity. A set of rate equations is derived, which agree very well with a solution of the full master equation model and makes it simple to investigate the properties of the system. Using a linearized version of these rate equations, we can analytically express the response of the nanolaser to a modulation of the pumping rate. These results are compared to the modulation response obtained directly from the master equation using a novel method. Using the rate equation method, we calculate the modulation bandwidth and show that, contrary to conventional semiconductor lasers, the nanolaser is typically over-damped and displays a dip in the modulation bandwidth as the two-level systems become inverted. Both these features can be traced back to the modeling of the emitters as two-level systems that are incoherently pumped.

\end{abstract}

\maketitle

The laser has evolved from table size apparatuses to truly nano sized devices, in much the same way that computer chips have been continuously minimized. The single or few-emitter nanolaser represents an extreme in terms of size. The emitters can be either atoms or quantum dots (QDs) that are coupled to a high-Q optical cavity, with mode volume on the order of the wavelength of light cubed. 
In these systems, the low number of emitters is compensated by a strong light-matter coupling and high Q values. Indeed, these nanolasers operate in a regime where experiments in cavity quantum electrodynamics (cQED) previously have shown strong-coupling behavior such as Rabi-splitting  \cite{Peter2005,Thompson1992,Takao2006}.

The experimental realization of such nanolasers is relatively new. For atoms the experiment may be performed by letting the atoms fall through the cavity and thereby obtaining a strong light-matter coupling during the transit period \cite{kimble2003}.
For lasers based on semiconductor QDs, the optical cavity is in the form of a planar photonic crystal \cite{nomura2010,strauf2006} or a micropillar \cite{reitzenstein2008}.
A strong coupling between the QDs and cavity is obtained either by defining the cavity around a preselected QD \cite{badolato2005,nomura2010}, or by selecting a device with good spatial and spectral match from many samples. 
Although such lasers are in their infancy, they may have a promising future in applications where their small size and low power consumption are important, e.g. for on-chip optical interconnects and photonic integration. Other examples of lasers with one or a few emitters include ions in optical traps \cite{Meyer1997} and artificial atoms in quantum circuits \cite{astafiev2007}.

In the simplest model for such a nanolaser, the emitters are modeled as two-level systems, which are coupled to a single mode of the electromagnetic field in an optical cavity. The two-level emitters are pumped incoherently, i.e. by using a reverse decay rate from the ground to the excited state. Such models have previously been investigated \cite{MuY1992,loffler1997,delvalle2009,delvalle2007,Gartner2011,Gartner2012,Auffeves2010,Auffeves2011,Laussy2012,ReviewStrauf} and enjoy popularity, especially because of their simplicity.

The optical cavity changes the local density of optical states experienced by the emitters. This means that light-matter interactions are suppressed for some frequencies and Purcell enhanced \cite{purcell1946,gregersen2012} for others, namely for those at the cavity resonance.
Such lasers may become thresholdless, meaning that upon pumping the laser will start lasing, without the abrupt change in output power and coherence known from conventional lasers \cite{Carmichael1994,ReviewStrauf}.
However, for an incoherently pumped two-level emitter, a large pumping rate will dephase the transition to such an extent that the output will eventually be quenched and the coherence will be lost \cite{MuY1992,loffler1997}. Thus, the output of a few-emitter nanolaser can be divided into three regimes: the low pump regime where the emission is either anti-bunched or strongly bunched depending on the number of emitter, the laser regime at intermediate pumping rates where the emission has a poissonian distribution, and a quenched regime with chaotic emission at larger pumping rates.

In this paper we derive a rate equation model for a nanolaser system with an arbitrary number of single emitters coupled to a cavity. 
This derivation is based on the Von Neumann master equation with Lindblad terms \cite{carmichael}, which is widely used in cQED models.
 The rate equations are easy to solve and give physical insight, as opposed to numerical solutions of the full master equation.  We proceed by using these rate equations to analytically investigate the response of a nanolaser when subject to a modulation of the pumping rate, and thereby gain a deeper insight into the fundamental dynamics of few-emitter nanolasers. Modulation responses have previously been investigated both experimentally \cite{altuh2006,englund2008}  and theoretically \cite{lorke2010,suhr2011} for lasers containing many emitters, but the present paper is the first to calculate these properties for system with only a few emitters. Previous investigations have predicted that nanolasers can exhibit a very large modulation bandwidth caused by the Purcell enhanced spontaneous emission rate \cite{altuh2006,suhr2011}. This effect should be even more pronounced in few emitter lasers, since they rely on a large Purcell enhancement to obtain lasing. 
We calculate the modulation bandwidth for the nanolaser and show that it displays a characteristic dip when the system is lasing. 
We also show that the response of the system to a modulation is over-damped, whereas conventional semiconductor lasers typically are under-damped.  The models developed in this paper can be used to assess the applicability of nanolasers in on-chip communication systems. 

\section{Model}
\begin{figure}[t]
\centering
\includegraphics[width=0.4\textwidth]{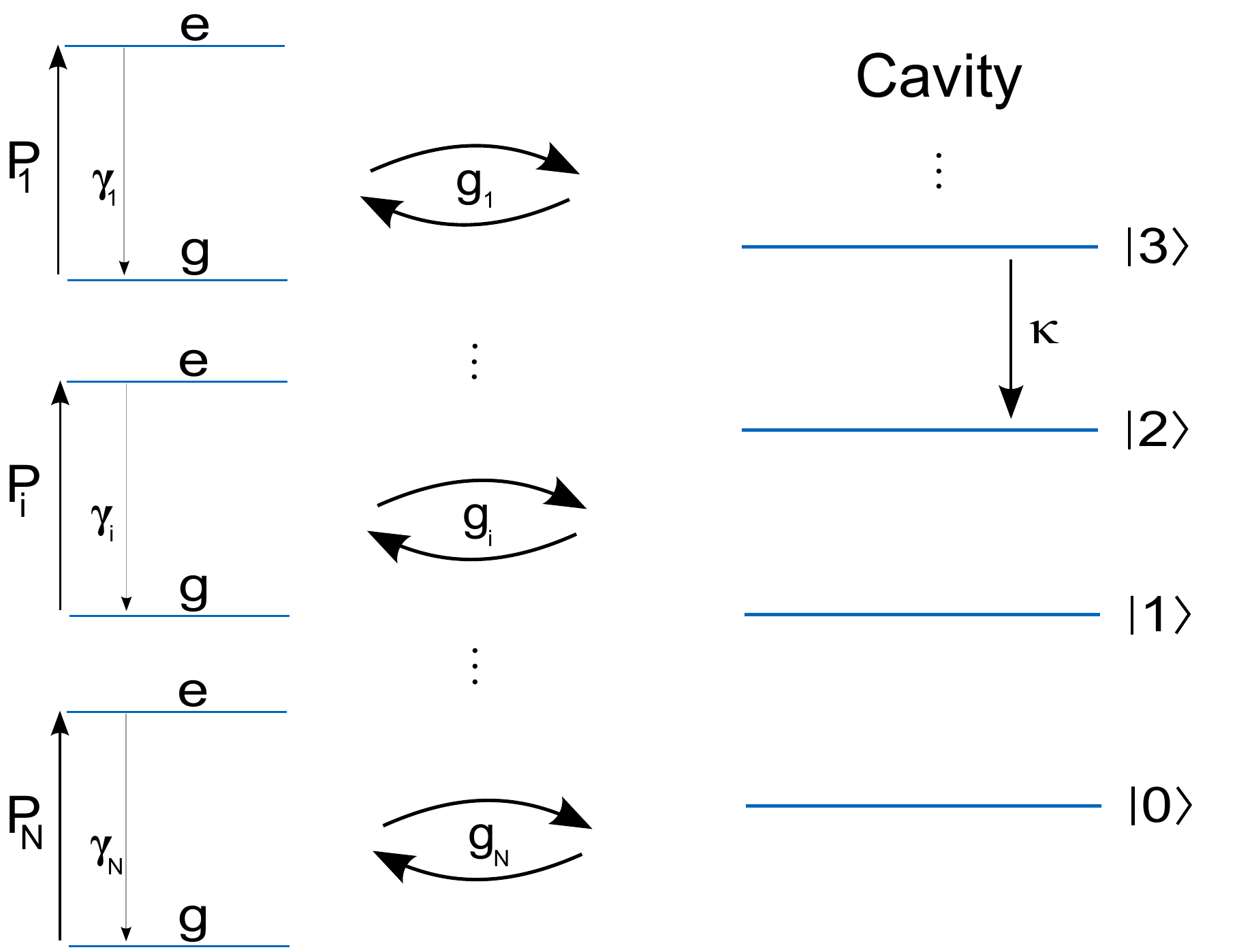}
\caption{\footnotesize{Schematic of the system considered where $N$ emitters are coupled to a single optical cavity with the coupling strengths $g_i$. The cavity decays at the rate $\kappa$ and the emitters are subject to a background decay rate $\gamma_i$ and an incoherent pumping rate $P_i$.}}
\label{TwoLevel}
\end{figure}

A two-level model with incoherent pumping is one of the simplest possible models for quantum emitters and has enjoyed large popularity for analysing cQED systems. It has therefore been used extensively to model single emitter lasers \cite{MuY1992,loffler1997,Jones1999,delvalle2009,Auffeves2010,Gartner2011,Gartner2012,Laussy2012}. Incoherent pumping models the excitation of an electron from the ground to the excited state, as e.g. from one atomic shell to another. In a semiconductor perspective this can be seen as a correlated creation of a hole in the valence band and an electron in the conduction band, i.e. the creation of an exciton.
 Although this model has also been used previously for modeling laser based on semiconductor QDs \cite{delvalle2007,delvalle2009}, it lacks several features which are relevant in QDs. For a more realistic model of QDs, it may be important to include additional levels with a more sophisticated pumping method and effects such as Coulomb interaction \cite{Gies1997,Ritter2010,GiesPSS2011,Gies2011} and hybridization with states of the wetting layer \cite{Imamoglu2007,Finley2008}. However, a two-level model can be used to model atomic system which are generally free of the environmental effect of semiconductors, and may also be used to model lasing in systems consisting of superconducting qubits \cite{Cole2010,astafiev2007}. Nevertheless, the general features that we predict and analyze should be relevant also for QD based lasers.

The nanolasers considered here contain $N$ emitters, modeled as two-level systems and coupled to a single mode of a cavity. See Fig. \ref{TwoLevel} for a schematic. The light-matter coupling strength between the $i$'th emitter and the single mode of the optical cavity is $g_i$, and the emitters are pumped individually with an incoherent pumping term with the rate $P_i$. Additionally, the emitters can decay with a rate $\gamma_i$,  which models spontaneous decay to other modes than the primary mode of the cavity, and also accounts for any non-radiative relaxation of the emitter. Finally, light is coupled out of the cavity at the rate $\kappa$. 

To model the different decay channels, the system is coupled to a number of reservoirs. These reservoirs are assumed to be time-independent and memory-less. In this case the Born-Markov approximation can be made and the resulting master equation for the combined emitter-cavity system is \cite{carmichael}
\begin{equation} \label{mastereq}
	\frac{d\rho(t)}{dt} = \frac{1}{i\hbar}\left[ H_{\rm JC},\rho(t)\right] - \mathcal{L}_P\rho(t) - \mathcal{L}_\gamma\rho(t) -\mathcal{L}_\kappa\rho(t),
\end{equation}
where $\rho(t)$ is the density matrix. The Hamiltonian is of the Jaynes-Cummings form
\begin{equation}
 H_{\rm JC}=  \hbar\omega_c a^\dagger a + \sum_{i=1}^N \hbar\omega_i \sigma^\dagger_i \sigma_i + \sum_{i=1}^N \hbar g_i \left( \sigma_i a^\dagger + \sigma_i^\dagger a \right),
\end{equation}
where $\sigma^\dagger_i$ ($\sigma_i$) is the raising (lowering) operator for the $i$'th emitter, which has frequency $\omega_i$, and where $a$ and $a^\dagger$ are the creation and annihilation operators for the single photonic mode of the cavity with frequency $\omega_c$. We shall assume $\omega_i=\omega_c$, corresponding to zero detuning between the emitters and the cavity. For identical emitters it is possible to tune the emitters and cavity into resonance. In the case of inhomogeneous broadening, such as for Stranski-Krastanov grown QDs, the cavity may act to select those QDs that are in resonance, but a finite detuning may need to be included.

The Lindblad terms in the master equation are
\begin{equation}
   \mathcal{L}_\kappa\rho(t) = \frac{\kappa}{2} \left(a^\dagger a\rho(t) + \rho(t)a^\dagger a - 2a\rho(t)a^\dagger\right)
\end{equation}
for the cavity decay and
\begin{eqnarray}
   \mathcal{L}_x\rho(t) &=& \sum_{i=1}^N \frac{x^{(i)}}{2} \left({\Gamma^{(i)}_x}^\dagger \Gamma^{(i)}_x\rho(t) + \rho(t){\Gamma^{(i)}_x}^\dagger \Gamma^{(i)}_x \right. \nonumber\\
   && \left. - 2\Gamma^{(i)}_x\rho(t){\Gamma^{(i)}_x}^\dagger\right),
\end{eqnarray}
where $\mathcal{L}_P\rho(t)$, with $\Gamma^{(i)}_P = \sigma_i^\dagger$, represents the pumping process where the emitter is excited from the ground to the excited state. The background decay, which includes spontaneous emission to other modes than the cavity mode and non-radiative recombination, is represented by $\mathcal{L}_\gamma\rho(t)$ with $\Gamma^{(i)}_\gamma =  \sigma_i$.

From Eq. (\ref{mastereq}) it is easily shown that the time evolution of the expectation value of a general operator $A$ is \cite{Gartner2012}
\begin{align}\label{avA}
   \frac{d \average{A}}{dt} =& \frac{d}{dt} {\rm Tr} \left[ A \rho(t) \right]  \nonumber \\ 
   =& \frac{1}{i\hbar} \average{\left[A,H_{\rm JC} \right]} + \frac{\kappa}{2}\average{\left[a^\dagger,A\right]a+a^\dagger\left[A,a\right]} \nonumber \\ 
   +&\sum_{x\in\{ P,\gamma\}} \sum_{i}^N \frac{x^{(i)}}{2}  \Bigl\langle \left[\Gamma_x^{(i)^\dagger},A\right]\Gamma_x^{(i)}  \nonumber \\ 
    + & \Gamma_x^{(i)^\dagger}\left[A,\Gamma_x^{(i)}\right] \Bigr\rangle.
\end{align}
It shows that the time-evolution of an expectation value $\average{A}$ is in general linked to the time-evolution of higher-order expectation values, such as e.g. $\average{A a^\dagger\sigma}$ and $\average{a^\dagger A a}$. This expression will be used to derive the rate equations in the next section.

\begin{figure}[t]
\centering
\includegraphics[width=0.5\textwidth]{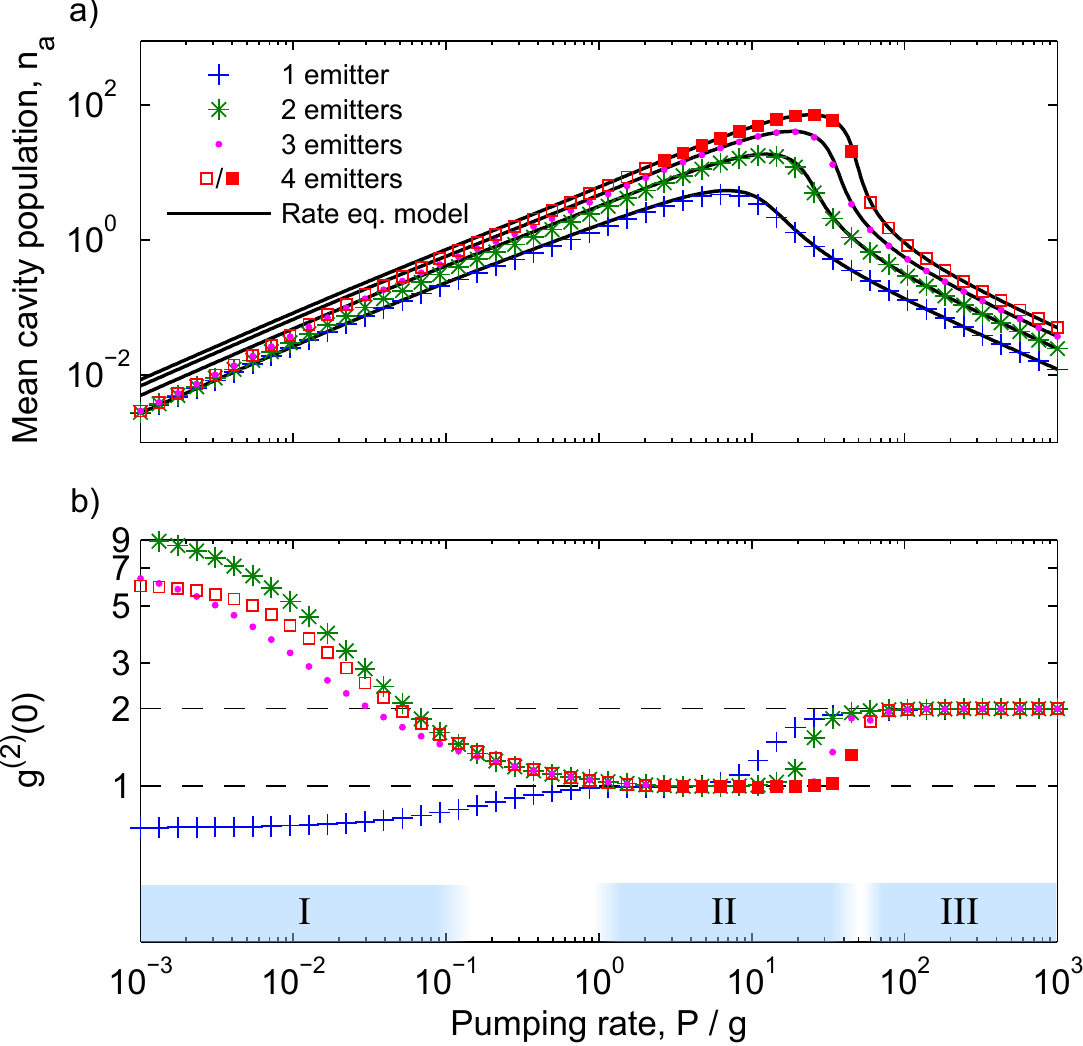}
\caption{\footnotesize{(color online) a) The mean cavity population as a function of the pumping rate for systems with between one and four emitters. The full lines are obtained using the rate equation model and the marks are numerical solutions to the master equation obtained using direct time-integration, except for the highest (marked with filled-in squares), which have been obtained using a Quantum Monte Carlo method. b) The calculated second order correlation function, $g^{(2)}(0)$ for the four cases. Three regimes are marked on the abscissa axis of b), corresponding to I: strongly bunched/anti-bunched emission, II: lasing and III: chaotic emission. The parameters are $g_i=g=300\;\rm ns^{-1}$ for the light-matter coupling, $\gamma_i=\gamma=10\;\rm ns^{-1}$ for the background decay rate and $\kappa=100\;\rm ns^{-1}$ for the cavity decay rate. }}
\label{NCavity}
\end{figure}

By using a basis consisting of the levels of the emitters and the number states of the cavity, the master equation can be transformed into a set of coupled linear differential equations. The Fock states of the cavity are truncated, so that only the lowest $N_c$ states are included. The choice of $N_c$ depends on, but is generally higher than, the mean population of the cavity. The resulting set of differential equations may then be solved using direct integration. However, when including several emitters and many states of the cavity, the Hilbert space grows rapidly, making direct integration unfeasible due to memory requirements. In this situation a Monte Carlo method \cite{carmichael1993,Molmer1993} may be employed, which trades run time for greatly reduced memory usage. For the Monte Carlo calculation we have used the software package QuTiP \cite{Johansson2012}.

In Fig. \ref{NCavity} a) the mean cavity population of systems with between one and four emitters is shown for different values of the pumping rate $P$. The Monte Carlo method has been used for the system with four emitters and high cavity population, and these results are shown with filled-in squares. For low pumping rates the cavity is scarcely populated (marked as regime I in the figure), but as the pumping rate is increased the mean number of photons in the cavity grows proportionally to the pumping rate $P$. This increase in the population of the cavity brings the system into the lasing regime, where the cavity photons stimulate the emission of further photons \cite{MuY1992} (regime II in the figure). Since the systems are nearly thresholdless, it is hard to distinguish regime I and II using the cavity populations alone. However, as will be discussed later, the statistics of the cavity field can be used to identify the regimes. For very large pumping rates the output quenches, i.e. the number of photons decreases towards zero, because the incoherent pumping dephases the transition from the emitter to the cavity (regime III in the figure). This phenomenon has been described before and is referred to as self-quenching \cite{MuY1992,loffler1997,Auffeves2010,Gartner2011}. These three regimes are general for systems with any number of emitters, as illustrated in Fig. \ref{NCavity} where the cavity populations for up to four emitters are shown. When more emitters are coupled to the cavity, the cavity population can grow larger before the quenching sets in. 

Reaching the lasing regime is conditioned on having a good cavity with a low cavity decay rate, $\kappa$, and a large light-matter coupling, $g$. For a system with $N$ identical emitters the collective light-matter coupling strength, $g_N=g\sqrt{N}$, increases with the number of emitters \cite{yamamoto}. Therefore, a system which cannot reach the lasing regime with a single emitter may be able to do so if more emitters are coupled to the cavity. 
 
Laser light is conventionally associated with a coherent state of the optical field, however it is well known that a coherent state is difficult, if not impossible, to obtain in quantum optics \cite{moelmer1997}. It can be shown that the cavity state of the systems described here can never be fully coherent by writing the equation of motion for $\average{a}$ using Eq. (\ref{avA}). This shows that $\average{a}$ depends only on $\average{\sigma}$ which in turn couples to higher order operator averages such as $\average{\sigma\sigma^\dagger a}$. But there are no direct driving terms for $\average{a}$, $\average{\sigma}$ or any of the other operator averages in the hierarchy, which means that in steady state $\average{a}$ is zero.  A coherent state, on the other hand, always has $|\average{a}|=\sqrt{\average{a^\dagger a}}$. 
 This conclusion has been reached previously by examining the structure of the master equation in Eq. (\ref{mastereq}), where it can be seen that there is no coupling between different Fock state of the cavity when the emitters states are identical \cite{delvalle2007}. This means that the reduced density  matrix for the cavity states, obtained by tracing out the degrees of freedom associated with the emitters, $\rho_{\rm cav}={\rm Tr_{emit}}[\rho]$,  only has diagonal elements.  This is different from the density matrix of a coherent state \cite{laudon}
\begin{equation}
\rho_{\rm coh} = e^{-|\alpha|^2} \sum_{n,m} \frac{\alpha^n\alpha^{*m} }{ \sqrt{n!m!}} |n\rangle 
\langle m|,
\end{equation}
which, has non-zero off-diagonal elements.
However, the system can be in a state where the diagonal terms of the density matrix are similar to those of a coherent state, but the others are zero \cite{delvalle2007}. Such a state is characterized by poissonian photon statistics. 

Formally the photon correlation functions of all orders are required to fully identify the photon statistics of the cavity field as poissonian, but the second order correlation is a good indicator of the statistics \cite{elvira2011}. The second order correlation function is defined as, $g^{(2)}(0)=\left(\average{n^2}-\average{n}\right)/\average{n}^2$. When $g^{(2)}(0)<1$ it indicates that the emission is anti-bunched and when $g^{(2)}(0)=2$ it signifies chaotic light, similar to that emitted by an incandescent light bulb. A poissonian photon distribution is characterized by \cite{laudon} $g^{(2)}(0)=1$, which is thus the criterion we use for lasing.

 In Fig. \ref{NCavity} b) the second order correlation is shown, and its value has been used to identify the three different regimes shown at the bottom of the figure. We see that in the low pump limit (regime I in the figure) the single emitter laser has a anti-bunched cavity field, and thus works as a single photon source. The cause of the anti-bunching of the emission is that the single emitter must be reexcited between two consecutive photon emissions, thus the probability of having a two-photon state in the cavity is small. Conversely, a system with more than one emitter can emit two photons into the cavity at the same time, and therefore shows a strong bunching behavior. We see that the system with three emitters shows a smaller bunching than the systems with two and four emitters. This dependency on the parity of the number of emitters is a sign of cooperativity, which means that the bunching is not accidental, but an indication that the emitters interact through the cavity field \cite{Auffeves2011}.
As the pumping rate is increased the emission of all the systems tends towards $g^{(2)}(0)=1$ representing lasing (regime II in the figure), but without showing a clear kink in the input-output curves, which is normally used as the criterion for reaching the threshold. As described previously, this is well known for lasers where a large fraction of the spontaneous emission goes into the lasing mode \cite{Carmichael1994,ReviewStrauf}.
As the pumping rate is increased further the systems experience self-quenching (regime III in the figure), as described above, and the cavity field becomes chaotic with $g^{(2)}(0)=2$.

\section{The rate equation approximation}\label{secrate}
Rate equation based models are very well established for modeling of conventional semiconductor lasers \cite{Coldren,Yamamoto1991}. However, conventional rate equations cannot immediately be applied to few emitter nanolasers. Indeed, a single emitter that is strongly coupled to a high quality cavity, i.e. with similar parameters as required for lasing, is known to operate in the regime of coherent Rabi oscillations. Such coherent oscillation usually cannot be predicted by a rate equation formalism. However, as we shall show, the large amount of dephasing which is induced by the incoherent pumping destroys the coherence and therefore enables a rate equation model description.

In deriving a set of rate equations, we start by writing the equation for the mean cavity population using Eq. (\ref{avA}) with $A=a^\dagger a$:
\begin{align}\label{eq_na}
   \frac{d}{dt} \average{a^\dagger a} &= \frac{d}{dt} \rm Tr \left[ a^\dagger a\rho(t)  \right]  \nonumber \\ 
   &= \sum_{i=1}^N 2 g_i {\rm Im} \left[ \average{\sigma_i a^\dagger}\right] - \kappa \average{a^\dagger a} .
\end{align}

Similarly we write the excited state probability of the $i$'th emitter as
\begin{align}\label{eq_ne}
   \frac{d}{dt} \average{\sigma_i^\dagger \sigma_i} &= \frac{d}{dt}  {\rm Tr}  \left[ \sigma_i^\dagger \sigma_i \rho(t)\right] \nonumber \\
   = &P_i \average{\sigma_i \sigma_i^\dagger}  -  2 g_i {\rm Im} \left[\average{\sigma_i a^\dagger}\right] - \gamma_i \average{\sigma_i^\dagger \sigma_i}.
\end{align}

We see that both $\average{a^\dagger a}$ and $\average{\sigma_i^\dagger \sigma_i}$ depend on the photon assisted polarization $\average{\sigma_i a^\dagger}$, which likewise can be derived from Eq. (\ref{avA}) as
\begin{align}\label{sigmaa}
   \left(\frac{d}{dt} + \frac{1}{2}(P_i+\gamma_i+\kappa) \right)\average{\sigma_i a^\dagger} = \nonumber \\
    i g_i \left( \average{\sigma_i^\dagger \sigma_i a a^\dagger} - \average{\sigma_i \sigma_i^\dagger a^\dagger a} \right)
\end{align}

At this point we make two approximations. 1)  we set $\frac{d}{dt}\average{\sigma_i a^\dagger}=0$, which means that the photon assisted polarization adiabatically follows the higher order operator averages $ \average{\sigma_i^\dagger \sigma_i a a^\dagger}$ and $\average{\sigma_i \sigma_i^\dagger a^\dagger a}$ . This is a good approximation when the system is not operating in the coherent regime and the dephasing is large. 
2) Using cluster expansion \cite{Gies1997,fricke1996}, we set $\average{\sigma_i^\dagger \sigma_i a a^\dagger} = \average{\sigma_i^\dagger \sigma_i} \average{a a^\dagger}+\average{\sigma_ia^\dagger }\average{\sigma_i^\dagger a}$ and $\average{\sigma_i \sigma_i^\dagger a^\dagger a}=\average{\sigma_i \sigma_i^\dagger}  \average{a^\dagger a}+\average{\sigma_ia^\dagger }\average{\sigma_i^\dagger a}$. The approximation consists of ignoring the quadruplet correlations \cite{Gies1997} $\delta\average{\sigma_i^\dagger \sigma_i a a^\dagger}$ and $\delta\average{\sigma_i\sigma_i^\dagger  a^\dagger a}$, which is justified when the dephasing of the photon assisted polarization, given by $\frac{1}{2}(P_i+\gamma_i+\kappa)$, is large. This specific application of the cluster expansion method is known as the doublet approximation \cite{Kira2006,Gies1997,fricke1996}. 

By solving for $\average{\sigma_i a^\dagger}$ in Eq. (\ref{sigmaa}) and inserting into Eq. (\ref{eq_na}) and (\ref{eq_ne}), we arrive at the rate equations
\begin{align}\label{rateeq_na}
   \frac{dn_a}{dt} &= \sum_{i=1}^N \left( F_i(n_e^{(i)}-n_g^{(i)})n_a + F_i n_e^{(i)} \right) - \kappa n_a \\ \label{rateeq_ne}
   \frac{dn_e^{(i)}}{dt} &= P_i n_g^{(i)} -F_i(n_e^{(i)}-n_g^{(i)})n_a - (F_i+\gamma_i) n_e^{(i)}
\end{align}
where $n_g^{(i)}=\average{\sigma_i \sigma_i^\dagger}$ and $n_e^{(i)}=\average{\sigma_i^\dagger \sigma_i}$ are the probabilities that the $i$'th emitter is in the ground or excited state respectively, and where $n_a=\average{a^\dagger a}$ is the mean cavity population. Since we assume that the electron is confined to the two-level emitter, it must be either in the ground or the excited state. Thus $n_e$ and $n_g$ are related by $n_g^{(i)} + n_e^{(i)} = 1$.  

The factors $F_i$, which enter the rate equation are given by
\begin{equation}\label{Frate}
 F_i = \frac{4g_i^2}{P_i + \gamma_i + \kappa}
\end{equation}
and are the Purcell enhanced rates into the cavity. They specify the coupling between the cavity and the $i$'th emitter. Importantly, we notice that they depend on the pumping rate, which enter the expression in a similar way as the decay rates.

The interpretation of the different terms in the rate equations is straight forward. The first term of Eq. (\ref{rateeq_na}), $F_i(n_e^{(i)}-n_g^{(i)})n_a$, corresponds to stimulated emission or absorption. If the emitter is not inverted, i.e. $n_e^{(i)}<n_g^{(i)}$, the term is negative which corresponds to absorption. Once the emitter is inverted the term becomes positive and corresponds to stimulated emission, which is responsible for the laser action. The second term $F_i n_e^{(i)}$ corresponds to spontaneous emission, which like the stimulated rate has been enhanced by the presence of the cavity \cite{gregersen2012}. The term $P n_g^{(i)}$ is the pumping term and depends on the occupation of the ground state.

The self-quenching seen in Fig. \ref{NCavity} can now be explained by considering $F_i$. The denominator in Eq. (\ref{Frate}) describes the dephasing of the transition, which becomes large, and is responsible for the decrease of the cavity population, as the pumping rate is increased. The mean number of photons in the cavity is a result of two competing processes. On one hand a higher pumping rate increases the excited state population of the emitters thereby increasing the stimulated emission into the cavity. On the other hand the increased pumping rate give rises to dephasing which decreases the coupling to the cavity. As the emitters become fully inverted, the stimulated emission rate cannot be increased further and the output quenches. In the limit of very large pumping rates, we see from Eq. (\ref{Frate}) that $F\rightarrow 0$ and the cavity population approaches zero \cite{MuY1992,loffler1997,delvalle2009,Gartner2011}.

The solution to the rate equations for one to five emitters is shown in Fig. \ref{NCavity} a). The results are compared with the values from the full model described in the previous section and, as can be seen, the two models agree well. In particular, the agreement is very good for large values of the pumping rate where, as discussed above, the large dephasing rate eliminates the higher order correlations between the emitter and the cavity, thus improving the approximations.

We note that from Eq. (\ref{eq_na}) and (\ref{eq_ne}) the following steady state relation can be derived 
\begin{equation}\label{naneP}
 \kappa n_a + \sum_{i=1}^N (P_i+\gamma_i)n_e^{(i)} =  \sum_{i=1}^N P_i.
\end{equation}

This relation has previously been derived for the one emitter case ($N=1$) \cite{delvalle2009,Gartner2011}, and is a very powerful relation which we will use later. It should also be noted that this expression is derived directly from the master equation, without the approximations used in deriving the rate equations.

\section{Modulation response}
The modulation response of a nanolaser can be calculated in several ways. One can directly induce a small periodic modulation of the pumping rate when the laser is in steady state, and simulate the response in dependence of the frequency. Alternatively, the linear response can be calculated directly from the equations governing the system. In this section we will use the rate equation to calculate the linear response of the system, and thereby obtain an analytical expression for the modulation bandwidth. In appendix \ref{app_linresp} we outline how to calculate the modulation response directly from the master equation and we use this method to validate the rate equation based method.

The behavior of the system is generally non-linear, but its small-signal response can be analyzed from a linearized version of the rate equations. They are linearized around the steady state values $\overline{n}_a$ and $\overline{n}_e$, by setting $n_a = \overline{n}_a + \delta n_a$ and $n_e^{(i)} = \overline{n}_e^{(i)} + \delta n_e^{(i)}$. The pumping term is expanded around the steady state value by writing it as $P_i = \overline{P_i} + \delta P$, and the rate $F_i$, which also depends on the pumping rate, as $F = \overline{F}_i + \frac{dF_i}{dP}\delta P=\overline{F}_i - \frac{\overline{F}_i^2}{4g^2}\delta P$. 
In matrix form the linearized rate equations for a single emitter ($N=1$) become
\begin{equation}\label{linrateeq_mx}
\frac{d}{dt} \left[ \begin{matrix}
\delta n_a  \\
\delta n_e  \end{matrix} \right] = \left[  \begin{matrix}
\gamma_{aa} & \gamma_{ae}  \\
-\gamma_{ea} & \gamma_{ee} \end{matrix} \right] \left[  \begin{matrix}
\delta n_a   \\
\delta n_e   \end{matrix} \right] + \delta P \left[  \begin{matrix}
- \gamma_{P}   \\
\overline{n}_g + \gamma_{P}   \end{matrix} \right].
\end{equation}
This is a linear system driven by the small perturbation, $\delta P$, in the pumping rate. The element $\gamma_{ae} = \overline{F}(2\overline{n}_a+1)$ determines the response of the cavity population to a variation of the emitter occupation. The first term originates from stimulated emission/absorption and the second term is the contribution from spontaneous emission. The other off-diagonal element, $\gamma_{ea} = \overline{F}(\overline{n}_e-\overline{n}_g)$, determines the response of the emitter caused by a small perturbation in the cavity population, and originates from the stimulated emission/absorption term in the rate equations. The diagonal elements are $\gamma_{aa} = \gamma_{ea} - \kappa$ and $\gamma_{ee} =  -\gamma_{ae}  - \gamma - \overline{P}$ and contain the same elements as the off-diagonal terms, in addition to terms originating from the decay of the cavity and emitters. Finally $\gamma_{P} =  \frac{\overline{F}^2}{4g^2}\left[(\overline{n}_e-\overline{n}_g) \overline{n}_a + \overline{n}_e \right]$ contains the contribution to the variation of $n_a$ and $n_e$ originating from the fact that $F$ varies with $P$.

With the linearized version of the rate equations it is easy to derive the modulation response. By taking the Fourier transform on both sides of Eq. (\ref{linrateeq_mx}) and solving for $\delta n_a(\omega)$, the modulation transfer function, $H(\omega) = \frac{\delta n_a(\omega)}{\delta n_a(0)}$, can be written as
\begin{equation}\label{modtransfer}
 H(\omega) = \frac{\left|  \begin{matrix}
\gamma_{aa} & \gamma_{ae}  \\
-\gamma_{ea} & \gamma_{ee}\end{matrix} \right| \left|  \begin{matrix}
-\gamma_{P}  & \gamma_{ae}  \\
\overline{n}_g + \gamma_{P}  & \gamma_{ee} +i\omega \end{matrix} \right|}{\left|  \begin{matrix}
\gamma_{aa} +i\omega & \gamma_{ae}  \\
-\gamma_{ea} & \gamma_{ee} +i\omega \end{matrix} \right|\left|  \begin{matrix}
-\gamma_{P}  & \gamma_{ae}  \\
\overline{n}_g + \gamma_{P}  & \gamma_{ee}\end{matrix} \right|}.
\end{equation}

\begin{figure}[tb!]
\centering
\includegraphics[width=0.5\textwidth]{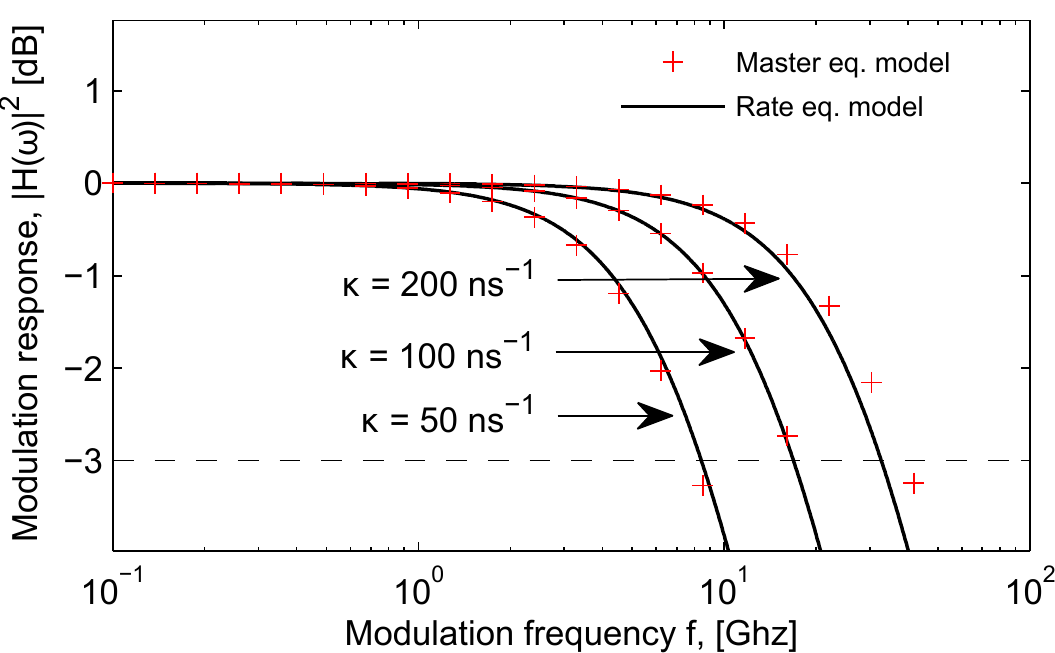}
\caption{\footnotesize{The modulation response $|H(\omega)|^2$ for a single emitter as a function of the frequency $f=\omega/2\pi$ for three different values of the cavity decay rate $\kappa$. The modulation response is calculated from the linearized rate equations in Eq. (\ref{modtransfer}) (full line) and directly from the master equation using the method outlined in appendix \ref{app_linresp} (marks). The horizontal line indicates a 3 dB decrease from the zero frequency response. The parameters are $P=g=300\;\rm ns^{-1}$ and $\gamma=10\;\rm ns^{-1}$.}}
\label{figmodulation}
\end{figure}

In Fig. \ref{figmodulation} the absolute square of the modulation transfer function $|H(\omega)|^2$, referred to as the modulation response, is shown for three different values of the cavity decay rate. The modulation response obtained using Eq. (\ref{modtransfer}) is compared with numerical results obtained directly from the master equation, as outlined in appendix \ref{app_linresp}. The rate equation model can be seen to agree quantitatively with the master equation results.

From Fig. \ref{figmodulation} it can also be seen that the modulation response decreases monotonically without the appearance of the usual relaxation oscillation resonance observed for conventional semiconductor lasers \cite{Coldren}. This can be investigated by calculating the discriminant for the characteristic equation of the matrix in Eq. (\ref{linrateeq_mx}) which is
\begin{eqnarray}\label{eqdisc}
 D = \left( \gamma_{ae} - \gamma_{ea}\right)^2 + 2\left( \Gamma-\kappa\right)\left(\gamma_{ae}+\gamma_{ea}\right) + \left(\Gamma-\kappa \right)^2,
\end{eqnarray}
where $\Gamma=\overline{P}+\gamma$. If the discriminant is positive the system is over-damped, and the solutions are exponentially decaying functions, whereas if $D$ is negative the systems is under-damped which means that the solutions are oscillatory, and a resonance may be seen in the modulation response. Since $\gamma_{ae} + \gamma_{ea}$ is positive, $D$ can only be negative if $\overline{P}<\kappa-\gamma$. Therefore, with the high pumping rate required for lasing, the system will always be over-damped while in the lasing regime.

The discriminant for a conventional semiconductor laser is similar to Eq. (\ref{eqdisc}), with the exception that $\Gamma$ is independent of the pumping rate. It is instead given by $\gamma_{\rm sp}$ which, similar to $\gamma$, is the spontaneous decay rate to other modes than the lasing mode and is independent of the pumping rate when the system is above the lasing threshold \cite{Coldren}. Since $\gamma_{\rm sp}$ typically is much smaller than $\kappa$, the discriminant can be negative, signifying an under-damped system. 

The main difference between the single emitter laser and the conventional semiconductor laser is thus the inclusion of the incoherent pumping rate $\overline{P}$ in the second term of the discriminant. This term stems from the way the pumping is represented in the model. In Eq. (\ref{rateeq_ne}) the pumping rate is $\overline{P} n_g$ which means that the pumping is dependent on the state of the emitter. This is contrary to a conventional semiconductor laser where the pumping term is just $P$. Using the relation $n_g+n_e=1$, we can write the pumping term as $\overline{P}(1-n_e)$, which gives the conventional population independent pumping rate, but also shows that there is an additional decay rate of the excited state, $-\overline{P} n_e$, as described above. This term enters in the same way as $\gamma n_e$ and therefore also gives rise to dephasing. The term is responsible for the $\overline{P}$-dependence of $\Gamma$ and is thus the origin of the fundamental difference in the dynamics between a conventional semiconductor laser and a nanolaser modeled as an incoherently pumped two-level system.

\section{The modulation bandwidth}
Figure \ref{figmodulation} shows that the intensity of the single emitter laser can follow the modulation until a certain frequency after which the response decreases drastically. The 3 dB frequency is the frequency at which the modulation response has decreased to half of its initial value and is also denoted the modulation bandwidth. In Fig. \ref{figmodulation} we see that as $\kappa$ is increased the modulation bandwidth also increases and in this section we will analyze the different parameters that influence the modulation bandwidth.

\begin{figure}[tb!]
\centering
\includegraphics[width=0.5\textwidth]{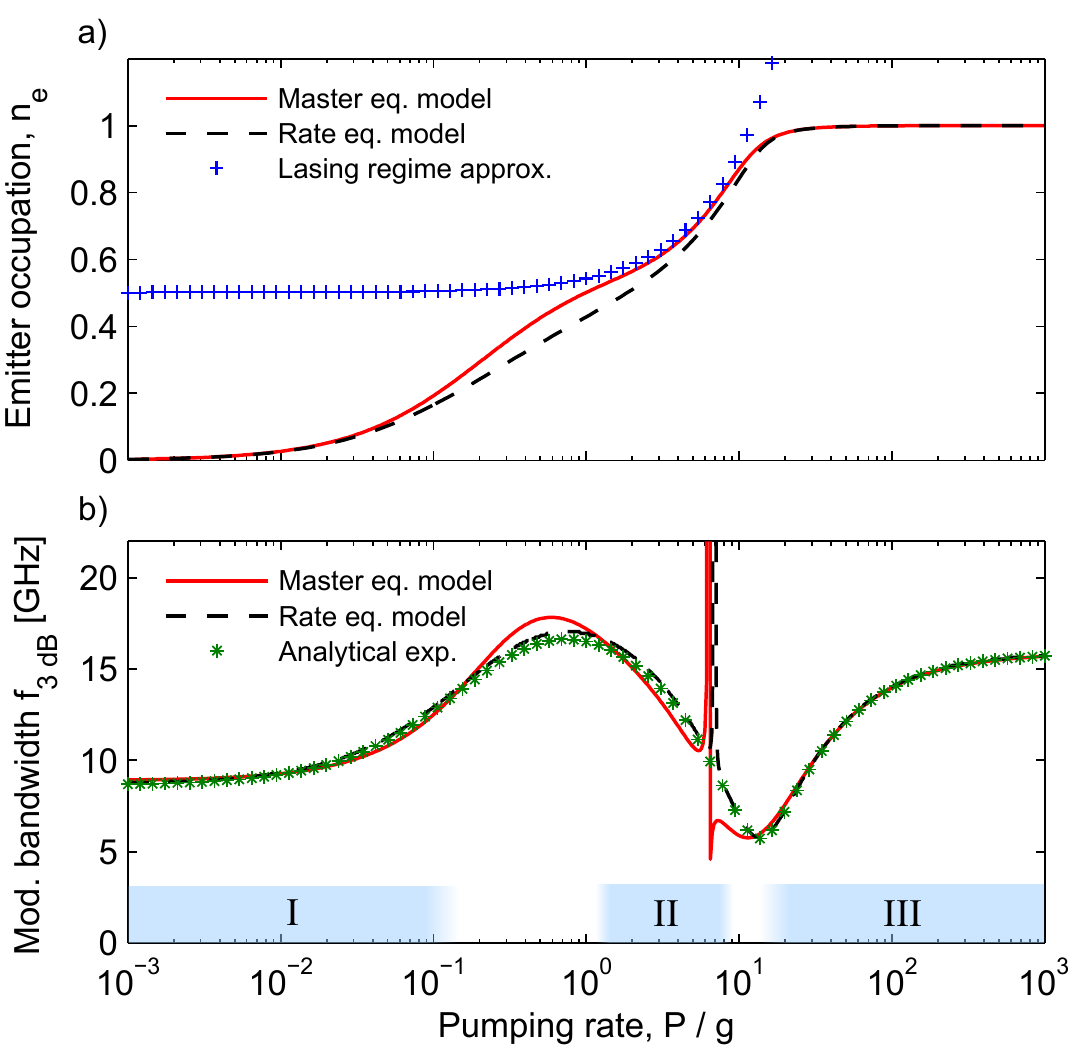}
\caption{\footnotesize{(color online) a) The probability that the excited state of the emitter is occupied, $n_e$, as a function of the pumping rate. The solution using the master equation (solid line) and the rate equations (dashed line) is shown, as well as the analytical approximation (marks), Eq. (\ref{ne1d}), which is valid in the lasing regime. b) The modulation bandwidth as a function of the pumping rate. The modulation bandwidth is found using the modulation response $H(\omega)$ as deduced from the master equation results using the method outlined in appendix \ref{app_linresp} (full line) or from the rate equations model using Eq. (\ref{modtransfer}) (dashed line). Also shown is the analytical expression from Eq. (\ref{3dbfreq}) (marks) . The parameters are $g=300\;\rm ns^{-1}$, $\kappa=100\;\rm ns^{-1}$ and $\gamma=10\;\rm ns^{-1}$. }}
\label{fig3db}
\end{figure}

In Fig. \ref{fig3db} b) we show the modulation bandwidth as a function of the pumping rate. The numerically exact modulation bandwidth obtained using the master equation, as described in appendix \ref{app_linresp}, is compared to the bandwidth found using the transfer function in Eq. (\ref{modtransfer}) which is based on the rate equations. The figure shows the same three regimes as Fig. \ref{NCavity}. We see that the rate equation model agrees very well with the master equation based model, except for a slight underestimation of the modulation bandwidth at intermediate values of the pumping rate. However, both models show a singularity in the modulation bandwidth at a certain value of the pumping rate in the lasing regime. The position of this singularity corresponds to the pumping rate at which the cavity population is at its maximum (see fig. \ref{NCavity}). At this point a small increase of the pumping rate, will not change the population in the cavity since the slope of $n_a$ is zero. Thus, the output modulation amplitude is zero, and the bandwidth is ill-defined.
It should be noted that this specific pumping rate also indicates the point where an increase of the pumping rate goes from giving a positive change of the cavity population to a negative change, as evidenced by the slope of the cavity population. 

The constant $\gamma_P$ in the expression for the modulation transfer function, Eq. (\ref{modtransfer}), describes the variation of $F$ when the pumping rate is changed. This variation is small and $\gamma_P=0$ can be assumed with only a minor change to the result. The modulation transfer function then becomes

\begin{equation}
 H(\omega) = \frac{\omega_R^2}{\omega_R^2-\omega^2+i\omega\gamma_m},
\end{equation}
where $\omega_R^2 = \gamma_{aa} \gamma_{ee}+\gamma_{ea} \gamma_{ae} $ is the determinant and $\gamma_m=\gamma_{ee} + \gamma_{aa}$ is the trace of the matrix in Eq. (\ref{linrateeq_mx}). Since we have ignored the pump dependence on $F$, this expression is similar to that of a conventional laser \cite{Coldren}.
Setting $|H(\omega_{\rm 3dB})|^2=\frac{1}{2}$, corresponding to a 3 dB decrease of the response, gives the modulation bandwidth
\begin{equation}\label{f3dbfull}
f_{\rm 3dB} = \frac{1}{2\pi}\sqrt{\omega_P^2 + \sqrt{\omega_P^4+\omega_R^4}},
\end{equation}
where $\omega_P^2 = \omega_R^2-\frac{1}{2}\gamma_m^2$. In the limit where $\gamma_m^2 \gg \omega_R^2$ a second order Taylor expansion of Eq. (\ref{f3dbfull}) in the parameter $\frac{\omega_R}{\gamma_m^2}$ gives
\begin{equation} \label{3dbfreq}
f_{\rm 3dB} = -\frac{1}{2\pi} \frac{\omega_R^2}{\gamma_m} =  -\frac{1}{2\pi} \frac{\gamma_{aa} \gamma_{ee}+\gamma_{ea} \gamma_{ae} }{\gamma_{ee} + \gamma_{aa}},
\end{equation}
which is the final analytical expression for the modulation bandwidth.
The discriminant in Eq. (\ref{eqdisc}) can be written $D=\gamma_m^2-4\omega_R^2$, and thus the criterion $\gamma_m^2 \gg \omega_R^2$ corresponds to an over-damped system. 
In this limit, the poles of the characteristic equation are 
\begin{equation} \label{poles}
s_1 = \frac{\omega_R^2}{\gamma_m} \hbox{\space\space and \space\space} s_2 = \gamma_m - \frac{\omega_R^2}{\gamma_m} \approx \gamma_m,
\end{equation}
where we see that $s_1$ exactly corresponds to the modulation bandwidth found above. This shows that the modulation bandwidth of the system is determined by the lower of the two poles of the system. These poles are different from those of a conventional semiconductor laser, where the poles normally form a conjugated pair of complex numbers \cite{Coldren}. By analyzing the modulation bandwidth it is thus possible to gain insight into the fundamental dynamic properties of the nanolaser.

In Fig. \ref{fig3db} b) the modulation bandwidth obtained with the approximate expression in Eq. (\ref{3dbfreq}) is compared with the numerically exact result found using the master equation model, as well as modulation bandwidth found using Eq. (\ref{modtransfer}). It can be seen that the two rate equation based results are very similar, justifying the approximation of disregarding $\gamma_P$.
Using the approximate formula we can analyze the modulation bandwidth in the different regimes; low, intermediate and high pumping rate. 

\subsection{Low pumping rate}\label{lowP}
In the limit of low pumping rate (regime I in Fig. \ref{fig3db}), we approximate $n_a\approx n_e\approx 0$, and the modulation bandwidth becomes 
\begin{equation}
	f_{\rm 3dB} \approx \frac{1}{2\pi}\frac{4g^2(\kappa+\gamma) + \kappa\gamma(\kappa+\gamma)}{8g^2+(\kappa+\gamma)^2},
\end{equation}
using Eq. (\ref{3dbfreq}).
In the limit where $\kappa\rightarrow0$ and $g\gg\gamma$ we have $f_{\rm 3dB}\rightarrow\frac{1}{4\pi }\gamma$. Likewise, when $\gamma\rightarrow0$ and $g\gg\kappa$ we have $f_{\rm 3dB}\rightarrow\frac{1}{4\pi }\kappa$. And for $\kappa\rightarrow\infty$ ($\gamma\rightarrow\infty$) we have  $f_{\rm 3dB}\rightarrow \frac{\gamma}{2\pi}$ ($f_{\rm 3dB}\rightarrow\frac{\kappa}{2\pi}$) as $\overline{P}\rightarrow0$. This shows that an excess cavity population $\delta n_a$ can decay through two channels; either directly out of the cavity at rate $\kappa$, or when the emitter and cavity are strongly coupled, it can reexcite the emitter which can then lose the excitation trough spontaneous emission at the rate $\gamma$. In this respect, the emitter effectively functions as a drain for the cavity population.

\subsection{Lasing}
When the system is lasing (regime II in Fig. \ref{fig3db}) the emitter is inverted, meaning that the probability of the emitter being in the excited state is larger than the probability of it being in the ground state, i.e. $n_e>n_g$. In Fig. \ref{fig3db} b) we see that the modulation bandwidth decreases as the emitter becomes increasingly inverted.
We can examine this using Eq. (\ref{3dbfreq}). $\gamma_m$ does not change much in this regime, so we can concentrate on the determinant $\omega_R^2$. Inserting $\gamma_{aa}$ and $\gamma_{ee}$ and using Eq. (\ref{naneP}) the determinant can be written
\begin{equation}
	\omega_R^2 = F\Big( \overline{P}-\gamma+\kappa + \Gamma\left(n_g-n_e\right) \Big) + \kappa\Gamma.
\end{equation}
The last term inside the parentheses becomes negative when the emitter becomes inverted, and is thus responsible for the decrease in the modulation bandwidth.
This effect would not be seen in conventional semiconductor lasers where, as discussed above, $\Gamma$ is independent of $P$ and small compared to the other rates. Thus we again see that the inclusion of the pumping rate in $\Gamma$ has a large impact on the dynamics of the system.

 The solution to the rate equations Eq. (\ref{rateeq_na}) - (\ref{rateeq_ne}) can be found analytically in the limit of small $\kappa$. The cavity population is described by \cite{Gartner2011}
\begin{equation}\label{na1d}
	n_a = \frac{\overline{P}-\gamma}{2\kappa} - \frac{(\overline{P}+\gamma)^2}{8g^2},
\end{equation}
and the excited state occupation is
\begin{equation}\label{ne1d}
n_e = \frac{1}{2} + \frac{\kappa(\overline{P}+\gamma)}{8g^2}.
\end{equation}
In the lasing regime the probability of the emitters being in the excited state, grows linearly from $n_e=\frac{1}{2}$ as the pumping rate is increased.
This approximation is shown in Fig. \ref{fig3db} a).
Inserting Eq. (\ref{na1d}) and (\ref{ne1d}) into Eq. (\ref{3dbfreq}) the modulation bandwidth becomes
\begin{equation}\label{f3dblasing}
 f_{\rm 3dB} \approx \frac{\kappa}{2\pi}\frac{4g^2(\overline{P}-\gamma+\kappa) + \kappa\Gamma(3\kappa-\Gamma)}{4g^2(\overline{P}-\gamma+\kappa) + 2\kappa^2(\kappa+\Gamma)},
\end{equation}
where it is apparent that the modulation bandwidth is very dependent on the cavity decay rate. However, since $\gamma\ll\overline{P}$ in this regime, the modulation bandwidth is almost independent of the background decay rate.

\subsection{High pumping rate}\label{highP}
For very large pumping rates, corresponding to regime III in Fig. \ref{fig3db}, the emitter is fully inverted ($n_e=1$), but because of self-quenching the cavity is empty ($n_a=0$). The modulation bandwidth is therefore
\begin{equation}
	f_{\rm 3dB} \approx \frac{1}{2\pi}\frac{F(\kappa-\Gamma)+\kappa\Gamma}{\kappa+\Gamma},
\end{equation}
which tends to $f_{\rm 3dB} \approx \frac{\kappa}{2\pi}$ when $P$ is large.
We see that the expression does not depend on $\gamma$. Unlike in the low pump rate limit, the emitter is fully inverted and cannot absorb a photon from the cavity, thus the emitter cannot function as a decay path for the excess cavity population. This cavity population can therefore only decay out of the cavity at rate $\kappa$ to bring the system back into steady state.

\section{Modulation bandwidth for systems with many emitters}

So far, we have calculated the modulation response only for a system with a single emitter. However, it is straight forwards to adapt the rate equation based expression for the modulation response to take several emitters into account. We see from Eq. (\ref{rateeq_na}) that if the emitters are similar, the only change in the equation for $n_a$ when going from one to $N$ emitters is $F\rightarrow N F$. This corresponds to a change of the light-matter coupling strength from $g$ to $\sqrt{N}g$ as described previously.
It means that in the case of $N$-emitters, the elements of the matrix for the linearized rate equations are
\begin{eqnarray}
 \gamma_{ae} &=& N\overline{F}(2\overline{n}_a+1), \\
 \gamma_{ea} &=& \overline{F}(2\overline{n}_e-1), \\
 \gamma_{aa} &=& N\gamma_{ea} - \kappa,  \\
 \gamma_{ee} &=&  -\gamma_{ae}/N  - \gamma, - \overline{P}
\end{eqnarray}
which can be used directly in Eq. (\ref{3dbfreq}) to obtain the modulation bandwidth. 

\begin{figure}[t!]
\centering
\includegraphics[width=0.5\textwidth]{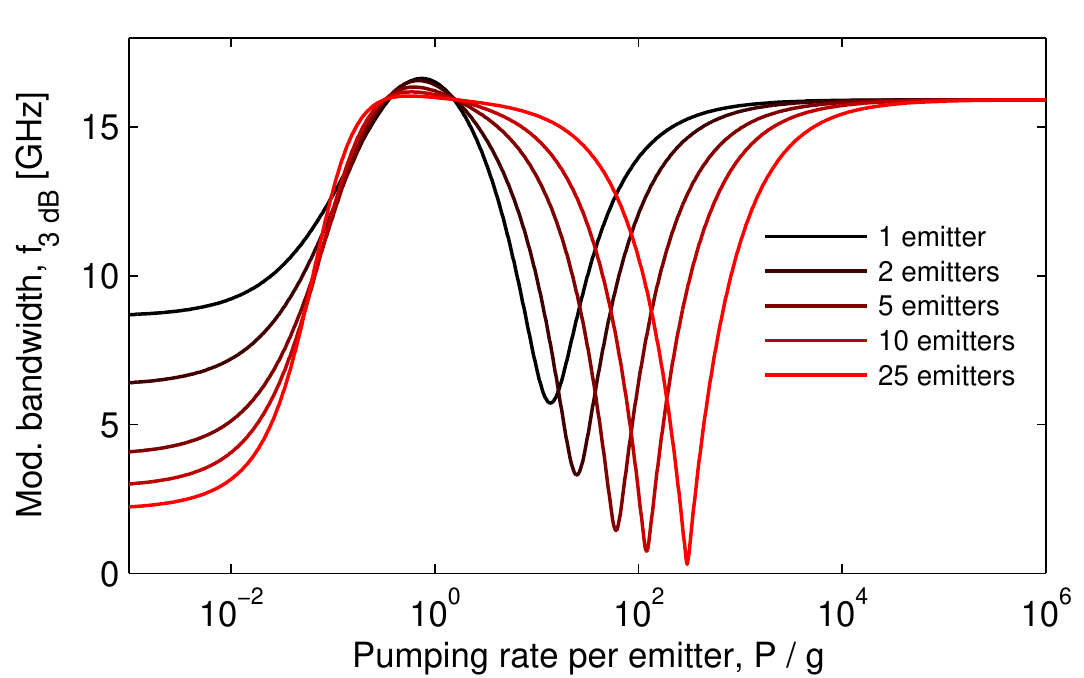}
\caption{\footnotesize{(color online) The modulation bandwidth as a function of the pumping rate per emitter for $N=\{1,2,5,10,25\}$ obtained using the rate equation model generalized to $N$-emitters. The parameters are $g_i=g=300\;\rm ns^{-1}$, $\kappa=100\;\rm ns^{-1}$ and $\gamma_i=\gamma=10\;\rm ns^{-1}$ }}
\label{fig3db_manydots}
\end{figure}

In Fig. \ref{fig3db_manydots} the modulation bandwidths as a function of the pumping rate per emitter for up to 25 emitters are shown. These results have been obtained using the rate equation based method generalized to $N$-emitters. In general we see that the form of the curves is similar to the one emitter case. The systems with many emitters show a larger dip in the modulation bandwidth, and in addition the pumping rate at which the emission quenches is larger, making the lasing regime broader. Also for more emitters, there seems to be a broader plateau of high modulation bandwidth before the dip, which indicates that these systems are less susceptible to the bandwidth lowering effect of the pumping method. However, the dips are deeper when the number of emitters is increased. The figure also shows that using more emitters does not increase the largest possible modulation frequency, since it is limited by the cavity decay rate $\kappa$. However, with more emitters it is possible to obtain lasing in a system with a small cavity Q-value, i.e. larger $\kappa$, giving a corresponding larger modulation bandwidth.

A generalization of the results in section \ref{lowP} and \ref{highP}, shows that $f_{\rm 3dB}\rightarrow\frac{1}{N+1}\frac{\kappa}{2\pi}$, ($f_{\rm 3dB}\rightarrow\frac{N}{N+1}\frac{\gamma}{2\pi}$) as $\gamma\rightarrow0$ ($\kappa\rightarrow0$) in the low pumping rate regime. In Fig. \ref{fig3db_manydots} we recognise the $\frac{1}{N+1}$ behavior which decreases the modulation bandwidth in systems with more emitters. In the high pumping rate regime we have $f_{\rm 3dB}\rightarrow\frac{\kappa}{2\pi}$, which is similar to the one emitter case.

\section{Summary}
In this article, we have derived a set of rate equations, which can be used to model the behavior of nanolasers based on single or few two-level emitters. The steady state values of the cavity population and emitter occupation, have been compared with result obtained using a full master equation model. The models agree when the systems are in the lasing regime and when the systems are quenched at large pumping rates. 

Using these rate equations, we have calculated the dynamical properties of the nanolasers. We have shown that, contrary to conventional semiconductor lasers, these nanolasers typically respond in an over-damped way when subject to a modulation of the pumping rate. Using the rate equations, we have also been able to obtain analytical expression for the modulation bandwidth of the lasers, and shown they exhibit a non-trivial behavior, with e.g. a dip in the bandwidth for a pumping rate in the lasing regime. 
For comparison, we have calculated the modulation response and bandwidth from the more fundamental master equation using first order perturbation theory. These results are in good agreement with those obtained from the rate equations.

For systems with several emitters, we have seen the same behavior as the single emitter case, although the system can operate at higher pumping rates and sustain higher power levels, before the output quenches or the dip in the modulation bandwidth occurs.

In this work we have modeled the pumping process as an incoherent rate from the ground to the excited level of the emitter. We have seen that the pumping process has a large influence on the dynamics, and that because the pumping induces a large dephasing of the photon assisted polarization, it is possible to write a set of rate equation for the system.  The same dephasing was found to be responsible for the quenching of the output. We have also shown that the pumping rate enters the rate equations in a different way than in conventional semiconductor lasers, giving an additional decay of the excited state of the emitter. This extra term in the rate equations is responsible for a dip in the modulation bandwidth and the over-damping of the response.

\section{Acknowledgments}
The authors acknowledge financial support from Villum Fonden via the NAnophotonics for TErabit Communications (NATEC) center.

\appendix
\section{Calculating modulation response using perturbation theory}\label{app_linresp}
This appendix outlines how to calculate the modulation response directly from the fundamental master equation using first order perturbation theory.

The master equation, Eq. (\ref{mastereq}), for the density matrix is linear and can thus be written as 
\begin{equation}
\frac{d\vec{\rho}}{dt} = \mathbb{M} \vec{\rho},
\end{equation}
where $\vec{\rho}$ is a vector containing the elements of the density matrix. The steady state solution, $\vec{\rho}_{ss}$, is given by $\mathbb{M} \vec{\rho}_{ss}=0$

When the pumping rate is time-dependent the system matrix, $\mathbb{M}$, becomes time-dependent. If the system is perturbed by a small change in the incoherent pumping rate, $P$, the matrix can be written as $\mathbb{M} + \alpha \mathbb{M}_1(t)$, where $\mathbb{M}$ is time-independent and $\alpha \mathbb{M}_1(t)$ specifies the small perturbation. Using a series expansion for $\vec{\rho}$
\begin{equation}
\vec{\rho}(t) = \sum_{n=0}^{\infty} \alpha^n \vec{\rho}^{(n)}(t),
\end{equation}
the zero order solution is given by
\begin{equation}
\frac{d\vec{\rho}^{(0)}}{dt} = \mathbb{M} \vec{\rho}^{(0)},
\end{equation}
and the first order solution can be found from
\begin{equation}
\frac{d\vec{\rho}^{(1)}}{dt} = \mathbb{M} \vec{\rho}^{(1)}(t) + \mathbb{M}_1(t) \vec{\rho}^{(0)}(t).
\end{equation}

By writing $(\frac{d}{dt} - \mathbb{M} )\vec{\rho}^{(1)}(t)= \mathbb{M}_1(t) \vec{\rho}^{(0)}(t)$ it is apparent that $\mathbb{M}_1(t) \vec{\rho}^{(0)}(t)$ is the driving term. Defining the Greens function as
\begin{equation}\label{greensfunc1}
\left(\frac{d}{dt} - \mathbb{M} \right)\mathbb{G} (t-t')= \mathbb{I} \delta(t-t'),
\end{equation}
the first order correction can be written
\begin{equation}\label{rho0dif}
	\vec{\rho}^{(1)}(t) = \int dt' \mathbb{G}(t-t') \mathbb{M}_1(t') \vec{\rho}^{(0)}(t').
\end{equation}

The Greens function is found by Fourier transforming both sides of (\ref{greensfunc1}), and is
\begin{equation}
\mathbb{G}(\omega) = -\frac{1}{i\omega\mathbb{I} + \mathbb{M}}.
\end{equation}

Identifying the right side of Eq. (\ref{rho0dif}) as a convolution, the Fourier transform of $\vec{\rho}^{(1)}$ is
\begin{equation}\label{RhoOneOmega}
	\vec{\rho}^{(1)}(\omega) = \mathbb{G}(\omega) \widehat{\mathbb{F}_t}\left(\mathbb{M}_1(t) \vec{\rho}^{(0)}(t)\right),
\end{equation}
where $\mathbb{F}_t$ is the operator for the Fourier transform with respect to t.
The cavity occupation is given by
\begin{eqnarray}
	n_{cav}(t) &=&  {\rm Tr} \left[ a^\dagger a \rho(t)\right] \nonumber\\&=&
	 \sum_{n} n \big( \left\langle g,n |\rho(t)| g,n \right\rangle +  \left\langle e,n|\rho(t)|e,n  \right\rangle \big) \nonumber\\&=& \sum_{n,\lambda\in\{e,g \}} n \rho_{n,\lambda}(t).
\end{eqnarray}
Thus the Fourier transform of the first order correction to the cavity occupation is 
\begin{equation}
n_{cav}^{(1)}(\omega) = \sum_{n,\lambda\in\{e,g \}} n \rho_{n,\lambda}^{(1)}(\omega).
\end{equation}

The matrix $\mathbb{M}_1(t)= \mathbb{X}_1 \delta(t-t_{ss})$, where $t_{ss}$ is the point in time after which the system is assumed to be in steady state, specifies the perturbation, and contains a delta function in the elements where $\mathbb{M}$ contains the pumping rate $P$. All other elements are zero. Thus the Fourier transform in (\ref{RhoOneOmega}) becomes $\mathbb{X}_1 \widehat{\mathbb{F}}_t\left(\vec{\rho}^{(0)}(t)\delta(t-t_{ss})\right) = \mathbb{X}_1 \vec{\rho}^{(0)}(t_{ss}) \exp(-i\omega t_{ss})$. Therefore, the modulation transfer function, defined as $H(\omega) =  \frac{n_{cav}^{(1)}(\omega)}{n_{cav}^{(1)}(0)}$, is 
\begin{equation}
H(\omega) =  \frac{  \sum_{n,\lambda\in\{e,g \}} n  \left[\mathbb{G}(\omega) \mathbb{X}_1 \vec{\rho}^{(0)}_{ss}e^{-i\omega t_{ss}} \right]_{n,\lambda}}{   \sum_{n,\lambda\in\{e,g \}} n \left[\mathbb{G}(0) \mathbb{X}_1 \vec{\rho}^{(0)}_{ss} \right]_{n,\lambda}}  .
\end{equation}


\begin{thebibliography}{47}
\expandafter\ifx\csname natexlab\endcsname\relax\def\natexlab#1{#1}\fi
\expandafter\ifx\csname bibnamefont\endcsname\relax
  \def\bibnamefont#1{#1}\fi
\expandafter\ifx\csname bibfnamefont\endcsname\relax
  \def\bibfnamefont#1{#1}\fi
\expandafter\ifx\csname citenamefont\endcsname\relax
  \def\citenamefont#1{#1}\fi
\expandafter\ifx\csname url\endcsname\relax
  \def\url#1{\texttt{#1}}\fi
\expandafter\ifx\csname urlprefix\endcsname\relax\def\urlprefix{URL }\fi
\providecommand{\bibinfo}[2]{#2}
\providecommand{\eprint}[2][]{\url{#2}}

\bibitem[{\citenamefont{Peter et~al.}(2005)\citenamefont{Peter, Senellart,
  Martrou, Lema\^itre, Hours, G\'erard, and Bloch}}]{Peter2005}
\bibinfo{author}{\bibfnamefont{E.}~\bibnamefont{Peter}},
  \bibinfo{author}{\bibfnamefont{P.}~\bibnamefont{Senellart}},
  \bibinfo{author}{\bibfnamefont{D.}~\bibnamefont{Martrou}},
  \bibinfo{author}{\bibfnamefont{A.}~\bibnamefont{Lema\^itre}},
  \bibinfo{author}{\bibfnamefont{J.}~\bibnamefont{Hours}},
  \bibinfo{author}{\bibfnamefont{J.~M.} \bibnamefont{G\'erard}},
  \bibnamefont{and} \bibinfo{author}{\bibfnamefont{J.}~\bibnamefont{Bloch}},
  \bibinfo{journal}{Phys. Rev. Lett.} \textbf{\bibinfo{volume}{95}},
  \bibinfo{pages}{067401} (\bibinfo{year}{2005}).

\bibitem[{\citenamefont{Thompson et~al.}(1992)\citenamefont{Thompson, Rempe,
  and Kimble}}]{Thompson1992}
\bibinfo{author}{\bibfnamefont{R.~J.} \bibnamefont{Thompson}},
  \bibinfo{author}{\bibfnamefont{G.}~\bibnamefont{Rempe}}, \bibnamefont{and}
  \bibinfo{author}{\bibfnamefont{H.~J.} \bibnamefont{Kimble}},
  \bibinfo{journal}{Phys. Rev. Lett.} \textbf{\bibinfo{volume}{68}},
  \bibinfo{pages}{1132} (\bibinfo{year}{1992}).

\bibitem[{\citenamefont{Aoki et~al.}(2006)\citenamefont{Aoki, Dayan,
  E~Wilcut~and, Parkins, Kippenberg, Vahala, and Kimble}}]{Takao2006}
\bibinfo{author}{\bibfnamefont{T.}~\bibnamefont{Aoki}},
  \bibinfo{author}{\bibfnamefont{B.}~\bibnamefont{Dayan}},
  \bibinfo{author}{\bibfnamefont{W.~P.~B.} \bibnamefont{E~Wilcut~and}},
  \bibinfo{author}{\bibfnamefont{A.~S.} \bibnamefont{Parkins}},
  \bibinfo{author}{\bibfnamefont{T.~J.} \bibnamefont{Kippenberg}},
  \bibinfo{author}{\bibfnamefont{K.~J.} \bibnamefont{Vahala}},
  \bibnamefont{and} \bibinfo{author}{\bibfnamefont{H.~J.}
  \bibnamefont{Kimble}}, \bibinfo{journal}{Nature}
  \textbf{\bibinfo{volume}{443}}, \bibinfo{pages}{671–4}
  (\bibinfo{year}{2006}).

\bibitem[{\citenamefont{McKeever et~al.}(2003)\citenamefont{McKeever, Boca,
  Boozer, Buck, and Kimble}}]{kimble2003}
\bibinfo{author}{\bibfnamefont{J.}~\bibnamefont{McKeever}},
  \bibinfo{author}{\bibfnamefont{A.}~\bibnamefont{Boca}},
  \bibinfo{author}{\bibfnamefont{A.~D.} \bibnamefont{Boozer}},
  \bibinfo{author}{\bibfnamefont{J.~R.} \bibnamefont{Buck}}, \bibnamefont{and}
  \bibinfo{author}{\bibfnamefont{H.~J.} \bibnamefont{Kimble}},
  \bibinfo{journal}{Nature} \textbf{\bibinfo{volume}{425}},
  \bibinfo{pages}{268} (\bibinfo{year}{2003}).

\bibitem[{\citenamefont{Nomura et~al.}(2010)\citenamefont{Nomura, Kumagai,
  Iwamoto, Ota, and Arakawa}}]{nomura2010}
\bibinfo{author}{\bibfnamefont{M.}~\bibnamefont{Nomura}},
  \bibinfo{author}{\bibfnamefont{N.}~\bibnamefont{Kumagai}},
  \bibinfo{author}{\bibfnamefont{S.}~\bibnamefont{Iwamoto}},
  \bibinfo{author}{\bibfnamefont{Y.}~\bibnamefont{Ota}}, \bibnamefont{and}
  \bibinfo{author}{\bibfnamefont{Y.}~\bibnamefont{Arakawa}},
  \bibinfo{journal}{Nature Phys} \textbf{\bibinfo{volume}{6}},
  \bibinfo{pages}{279 } (\bibinfo{year}{2010}).

\bibitem[{\citenamefont{Strauf et~al.}(2006)\citenamefont{Strauf, Hennessy,
  Rakher, Choi, Badolato, Andreani, Hu, Petroff, and Bouwmeester}}]{strauf2006}
\bibinfo{author}{\bibfnamefont{S.}~\bibnamefont{Strauf}},
  \bibinfo{author}{\bibfnamefont{K.}~\bibnamefont{Hennessy}},
  \bibinfo{author}{\bibfnamefont{M.~T.} \bibnamefont{Rakher}},
  \bibinfo{author}{\bibfnamefont{Y.-S.} \bibnamefont{Choi}},
  \bibinfo{author}{\bibfnamefont{A.}~\bibnamefont{Badolato}},
  \bibinfo{author}{\bibfnamefont{L.~C.} \bibnamefont{Andreani}},
  \bibinfo{author}{\bibfnamefont{E.~L.} \bibnamefont{Hu}},
  \bibinfo{author}{\bibfnamefont{P.~M.} \bibnamefont{Petroff}},
  \bibnamefont{and}
  \bibinfo{author}{\bibfnamefont{D.}~\bibnamefont{Bouwmeester}},
  \bibinfo{journal}{Phys. Rev. Lett.} \textbf{\bibinfo{volume}{96}},
  \bibinfo{pages}{127404} (\bibinfo{year}{2006}).

\bibitem[{\citenamefont{Reitzenstein et~al.}(2008)\citenamefont{Reitzenstein,
  B\"{o}ckler, Bazhenov, Gorbunov, L\"{o}ffler, Kamp, Kulakovskii, and
  Forchel}}]{reitzenstein2008}
\bibinfo{author}{\bibfnamefont{S.}~\bibnamefont{Reitzenstein}},
  \bibinfo{author}{\bibfnamefont{C.}~\bibnamefont{B\"{o}ckler}},
  \bibinfo{author}{\bibfnamefont{A.}~\bibnamefont{Bazhenov}},
  \bibinfo{author}{\bibfnamefont{A.}~\bibnamefont{Gorbunov}},
  \bibinfo{author}{\bibfnamefont{A.}~\bibnamefont{L\"{o}ffler}},
  \bibinfo{author}{\bibfnamefont{M.}~\bibnamefont{Kamp}},
  \bibinfo{author}{\bibfnamefont{V.~D.} \bibnamefont{Kulakovskii}},
  \bibnamefont{and} \bibinfo{author}{\bibfnamefont{A.}~\bibnamefont{Forchel}},
  \bibinfo{journal}{Opt. Express} \textbf{\bibinfo{volume}{16}},
  \bibinfo{pages}{4848} (\bibinfo{year}{2008}).

\bibitem[{\citenamefont{Badolato et~al.}(2005)\citenamefont{Badolato, Hennessy,
  Atatüre, Dreiser, Hu, Petroff, and Imamoglu}}]{badolato2005}
\bibinfo{author}{\bibfnamefont{A.}~\bibnamefont{Badolato}},
  \bibinfo{author}{\bibfnamefont{K.}~\bibnamefont{Hennessy}},
  \bibinfo{author}{\bibfnamefont{M.}~\bibnamefont{Atatüre}},
  \bibinfo{author}{\bibfnamefont{J.}~\bibnamefont{Dreiser}},
  \bibinfo{author}{\bibfnamefont{E.}~\bibnamefont{Hu}},
  \bibinfo{author}{\bibfnamefont{P.~M.} \bibnamefont{Petroff}},
  \bibnamefont{and} \bibinfo{author}{\bibfnamefont{A.}~\bibnamefont{Imamoglu}},
  \bibinfo{journal}{Science} \textbf{\bibinfo{volume}{308}},
  \bibinfo{pages}{1158} (\bibinfo{year}{2005}).

\bibitem[{\citenamefont{Meyer et~al.}(1997)\citenamefont{Meyer, Briegel, and
  Walther}}]{Meyer1997}
\bibinfo{author}{\bibfnamefont{G.~M.} \bibnamefont{Meyer}},
  \bibinfo{author}{\bibfnamefont{H.-J.} \bibnamefont{Briegel}},
  \bibnamefont{and} \bibinfo{author}{\bibfnamefont{H.}~\bibnamefont{Walther}},
  \bibinfo{journal}{EPL (Europhysics Letters)} \textbf{\bibinfo{volume}{37}},
  \bibinfo{pages}{317} (\bibinfo{year}{1997}).

\bibitem[{\citenamefont{Astafiev et~al.}(2007)\citenamefont{Astafiev, Inomata,
  Niskanen, Yamamoto, Pashkin, Nakamura, and Tsai}}]{astafiev2007}
\bibinfo{author}{\bibfnamefont{O.}~\bibnamefont{Astafiev}},
  \bibinfo{author}{\bibfnamefont{K.}~\bibnamefont{Inomata}},
  \bibinfo{author}{\bibfnamefont{A.~O.} \bibnamefont{Niskanen}},
  \bibinfo{author}{\bibfnamefont{T.}~\bibnamefont{Yamamoto}},
  \bibinfo{author}{\bibfnamefont{Y.~A.} \bibnamefont{Pashkin}},
  \bibinfo{author}{\bibfnamefont{Y.}~\bibnamefont{Nakamura}}, \bibnamefont{and}
  \bibinfo{author}{\bibfnamefont{J.~S.} \bibnamefont{Tsai}},
  \bibinfo{journal}{Nature} \textbf{\bibinfo{volume}{449}},
  \bibinfo{pages}{588} (\bibinfo{year}{2007}).

\bibitem[{\citenamefont{Mu and Savage}(1992)}]{MuY1992}
\bibinfo{author}{\bibfnamefont{Y.}~\bibnamefont{Mu}} \bibnamefont{and}
  \bibinfo{author}{\bibfnamefont{C.~M.} \bibnamefont{Savage}},
  \bibinfo{journal}{Phys. Rev. A} \textbf{\bibinfo{volume}{46}},
  \bibinfo{pages}{5944} (\bibinfo{year}{1992}).

\bibitem[{\citenamefont{L\"offler et~al.}(1997)\citenamefont{L\"offler, Meyer,
  and Walther}}]{loffler1997}
\bibinfo{author}{\bibfnamefont{M.}~\bibnamefont{L\"offler}},
  \bibinfo{author}{\bibfnamefont{G.~M.} \bibnamefont{Meyer}}, \bibnamefont{and}
  \bibinfo{author}{\bibfnamefont{H.}~\bibnamefont{Walther}},
  \bibinfo{journal}{Phys. Rev. A} \textbf{\bibinfo{volume}{55}},
  \bibinfo{pages}{3923} (\bibinfo{year}{1997}).

\bibitem[{\citenamefont{del Valle et~al.}(2009)\citenamefont{del Valle, Laussy,
  and Tejedor}}]{delvalle2009}
\bibinfo{author}{\bibfnamefont{E.}~\bibnamefont{del Valle}},
  \bibinfo{author}{\bibfnamefont{F.~P.} \bibnamefont{Laussy}},
  \bibnamefont{and} \bibinfo{author}{\bibfnamefont{C.}~\bibnamefont{Tejedor}},
  \bibinfo{journal}{Phys. Rev. B} \textbf{\bibinfo{volume}{79}},
  \bibinfo{pages}{235326} (\bibinfo{year}{2009}).

\bibitem[{\citenamefont{del Valle et~al.}(2007)\citenamefont{del Valle, Laussy,
  Troiani, and Tejedor}}]{delvalle2007}
\bibinfo{author}{\bibfnamefont{E.}~\bibnamefont{del Valle}},
  \bibinfo{author}{\bibfnamefont{F.~P.} \bibnamefont{Laussy}},
  \bibinfo{author}{\bibfnamefont{F.}~\bibnamefont{Troiani}}, \bibnamefont{and}
  \bibinfo{author}{\bibfnamefont{C.}~\bibnamefont{Tejedor}},
  \bibinfo{journal}{Phys. Rev. B} \textbf{\bibinfo{volume}{76}},
  \bibinfo{pages}{235317} (\bibinfo{year}{2007}).

\bibitem[{\citenamefont{Gartner}(2011)}]{Gartner2011}
\bibinfo{author}{\bibfnamefont{P.}~\bibnamefont{Gartner}},
  \bibinfo{journal}{Phys. Rev. A} \textbf{\bibinfo{volume}{84}},
  \bibinfo{pages}{053804} (\bibinfo{year}{2011}).

\bibitem[{\citenamefont{Gartner}(2012)}]{Gartner2012}
\bibinfo{author}{\bibfnamefont{P.}~\bibnamefont{Gartner}},
  \bibinfo{journal}{Proc. SPIE 8440, Quantum Optics II}
  (\bibinfo{year}{2012}).

\bibitem[{\citenamefont{Auff\`eves et~al.}(2010)\citenamefont{Auff\`eves,
  Gerace, G\'erard, Santos, Andreani, and Poizat}}]{Auffeves2010}
\bibinfo{author}{\bibfnamefont{A.}~\bibnamefont{Auff\`eves}},
  \bibinfo{author}{\bibfnamefont{D.}~\bibnamefont{Gerace}},
  \bibinfo{author}{\bibfnamefont{J.-M.} \bibnamefont{G\'erard}},
  \bibinfo{author}{\bibfnamefont{M.~F.} \bibnamefont{Santos}},
  \bibinfo{author}{\bibfnamefont{L.~C.} \bibnamefont{Andreani}},
  \bibnamefont{and} \bibinfo{author}{\bibfnamefont{J.-P.}
  \bibnamefont{Poizat}}, \bibinfo{journal}{Phys. Rev. B}
  \textbf{\bibinfo{volume}{81}}, \bibinfo{pages}{245419}
  (\bibinfo{year}{2010}).

\bibitem[{\citenamefont{Auff\`ves et~al.}(2011)\citenamefont{Auff\`ves, Gerace,
  Portolan, Drezet, and Santos}}]{Auffeves2011}
\bibinfo{author}{\bibfnamefont{A.}~\bibnamefont{Auff\`ves}},
  \bibinfo{author}{\bibfnamefont{D.}~\bibnamefont{Gerace}},
  \bibinfo{author}{\bibfnamefont{S.}~\bibnamefont{Portolan}},
  \bibinfo{author}{\bibfnamefont{A.}~\bibnamefont{Drezet}}, \bibnamefont{and}
  \bibinfo{author}{\bibfnamefont{M.~F.} \bibnamefont{Santos}},
  \bibinfo{journal}{New Journal of Physics} \textbf{\bibinfo{volume}{13}},
  \bibinfo{pages}{093020} (\bibinfo{year}{2011}).

\bibitem[{\citenamefont{Laussy and del Valle;}(2012)}]{Laussy2012}
\bibinfo{author}{\bibfnamefont{F.~P.} \bibnamefont{Laussy}} \bibnamefont{and}
  \bibinfo{author}{\bibfnamefont{E.}~\bibnamefont{del Valle;}},
  \bibinfo{journal}{Proceedings of SPIE} \textbf{\bibinfo{volume}{8255}}
  (\bibinfo{year}{2012}).

\bibitem[{\citenamefont{Strauf and Jahnke}(2011)}]{ReviewStrauf}
\bibinfo{author}{\bibfnamefont{S.}~\bibnamefont{Strauf}} \bibnamefont{and}
  \bibinfo{author}{\bibfnamefont{F.}~\bibnamefont{Jahnke}},
  \bibinfo{journal}{Laser \& Photonics Reviews} \textbf{\bibinfo{volume}{5}},
  \bibinfo{pages}{607} (\bibinfo{year}{2011}).

\bibitem[{\citenamefont{Purcell}(1946)}]{purcell1946}
\bibinfo{author}{\bibfnamefont{E.}~\bibnamefont{Purcell}},
  \bibinfo{journal}{Phys. Rev.} \textbf{\bibinfo{volume}{69}},
  \bibinfo{pages}{681} (\bibinfo{year}{1946}).

\bibitem[{\citenamefont{Gregersen et~al.}(2012)\citenamefont{Gregersen, Suhr,
  Lorke, and Mork}}]{gregersen2012}
\bibinfo{author}{\bibfnamefont{N.}~\bibnamefont{Gregersen}},
  \bibinfo{author}{\bibfnamefont{T.}~\bibnamefont{Suhr}},
  \bibinfo{author}{\bibfnamefont{M.}~\bibnamefont{Lorke}}, \bibnamefont{and}
  \bibinfo{author}{\bibfnamefont{J.}~\bibnamefont{Mork}},
  \bibinfo{journal}{Appl. Phys. Lett.} \textbf{\bibinfo{volume}{100}},
  \bibinfo{eid}{131107} (\bibinfo{year}{2012}).

\bibitem[{\citenamefont{Rice and Carmichael}(1994)}]{Carmichael1994}
\bibinfo{author}{\bibfnamefont{P.~R.} \bibnamefont{Rice}} \bibnamefont{and}
  \bibinfo{author}{\bibfnamefont{H.~J.} \bibnamefont{Carmichael}},
  \bibinfo{journal}{Phys. Rev. A} \textbf{\bibinfo{volume}{50}},
  \bibinfo{pages}{4318} (\bibinfo{year}{1994}).

\bibitem[{\citenamefont{Carmichael}(1999)}]{carmichael}
\bibinfo{author}{\bibfnamefont{H.~J.} \bibnamefont{Carmichael}},
  \emph{\bibinfo{title}{Statistical Methods in Quantum Optics}},
  vol.~\bibinfo{volume}{1} (\bibinfo{publisher}{Springer},
  \bibinfo{year}{1999}).

\bibitem[{\citenamefont{Altug et~al.}(2006)\citenamefont{Altug, Englund, and
  Vuckovic}}]{altuh2006}
\bibinfo{author}{\bibfnamefont{H.}~\bibnamefont{Altug}},
  \bibinfo{author}{\bibfnamefont{D.}~\bibnamefont{Englund}}, \bibnamefont{and}
  \bibinfo{author}{\bibfnamefont{J.}~\bibnamefont{Vuckovic}},
  \bibinfo{journal}{Nature Physics} \textbf{\bibinfo{volume}{2}},
  \bibinfo{pages}{484–488} (\bibinfo{year}{2006}).

\bibitem[{\citenamefont{Englund et~al.}(2008)\citenamefont{Englund, Altug,
  Ellis, and Vuckovic}}]{englund2008}
\bibinfo{author}{\bibfnamefont{D.}~\bibnamefont{Englund}},
  \bibinfo{author}{\bibfnamefont{H.}~\bibnamefont{Altug}},
  \bibinfo{author}{\bibfnamefont{B.}~\bibnamefont{Ellis}}, \bibnamefont{and}
  \bibinfo{author}{\bibfnamefont{J.}~\bibnamefont{Vuckovic}},
  \bibinfo{journal}{Laser \& Photonics Reviews} \textbf{\bibinfo{volume}{2}},
  \bibinfo{pages}{264} (\bibinfo{year}{2008}).

\bibitem[{\citenamefont{Lorke et~al.}(2010)\citenamefont{Lorke, Nielsen, and
  Mork}}]{lorke2010}
\bibinfo{author}{\bibfnamefont{M.}~\bibnamefont{Lorke}},
  \bibinfo{author}{\bibfnamefont{T.~R.} \bibnamefont{Nielsen}},
  \bibnamefont{and} \bibinfo{author}{\bibfnamefont{J.}~\bibnamefont{Mork}},
  \bibinfo{journal}{Appl. Phys. Lett.} \textbf{\bibinfo{volume}{97}},
  \bibinfo{eid}{211106} (\bibinfo{year}{2010}).

\bibitem[{\citenamefont{Suhr et~al.}(2011)\citenamefont{Suhr, Gregersen, Lorke,
  and Mork}}]{suhr2011}
\bibinfo{author}{\bibfnamefont{T.}~\bibnamefont{Suhr}},
  \bibinfo{author}{\bibfnamefont{N.}~\bibnamefont{Gregersen}},
  \bibinfo{author}{\bibfnamefont{M.}~\bibnamefont{Lorke}}, \bibnamefont{and}
  \bibinfo{author}{\bibfnamefont{J.}~\bibnamefont{Mork}},
  \bibinfo{journal}{Appl. Phys. Lett.} \textbf{\bibinfo{volume}{98}},
  \bibinfo{eid}{211109} (pages~\bibinfo{numpages}{3}) (\bibinfo{year}{2011}).

\bibitem[{\citenamefont{Jones et~al.}(1999)\citenamefont{Jones, Ghose, Clemens,
  Rice, and Pedrotti}}]{Jones1999}
\bibinfo{author}{\bibfnamefont{B.}~\bibnamefont{Jones}},
  \bibinfo{author}{\bibfnamefont{S.}~\bibnamefont{Ghose}},
  \bibinfo{author}{\bibfnamefont{J.~P.} \bibnamefont{Clemens}},
  \bibinfo{author}{\bibfnamefont{P.~R.} \bibnamefont{Rice}}, \bibnamefont{and}
  \bibinfo{author}{\bibfnamefont{L.~M.} \bibnamefont{Pedrotti}},
  \bibinfo{journal}{Phys. Rev. A} \textbf{\bibinfo{volume}{60}},
  \bibinfo{pages}{3267} (\bibinfo{year}{1999}).

\bibitem[{\citenamefont{Gies et~al.}(2007)\citenamefont{Gies, Wiersig, Lorke,
  and Jahnke}}]{Gies1997}
\bibinfo{author}{\bibfnamefont{C.}~\bibnamefont{Gies}},
  \bibinfo{author}{\bibfnamefont{J.}~\bibnamefont{Wiersig}},
  \bibinfo{author}{\bibfnamefont{M.}~\bibnamefont{Lorke}}, \bibnamefont{and}
  \bibinfo{author}{\bibfnamefont{F.}~\bibnamefont{Jahnke}},
  \bibinfo{journal}{Phys. Rev. A} \textbf{\bibinfo{volume}{75}},
  \bibinfo{pages}{013803} (\bibinfo{year}{2007}).

\bibitem[{\citenamefont{Ritter et~al.}(2010)\citenamefont{Ritter, Gartner,
  Gies, and Jahnke}}]{Ritter2010}
\bibinfo{author}{\bibfnamefont{S.}~\bibnamefont{Ritter}},
  \bibinfo{author}{\bibfnamefont{P.}~\bibnamefont{Gartner}},
  \bibinfo{author}{\bibfnamefont{C.}~\bibnamefont{Gies}}, \bibnamefont{and}
  \bibinfo{author}{\bibfnamefont{F.}~\bibnamefont{Jahnke}},
  \bibinfo{journal}{Opt. Express} \textbf{\bibinfo{volume}{18}},
  \bibinfo{pages}{9909} (\bibinfo{year}{2010}).

\bibitem[{\citenamefont{Gies et~al.}(2011{\natexlab{a}})\citenamefont{Gies,
  Florian, Gartner, and Jahnke}}]{GiesPSS2011}
\bibinfo{author}{\bibfnamefont{C.}~\bibnamefont{Gies}},
  \bibinfo{author}{\bibfnamefont{M.}~\bibnamefont{Florian}},
  \bibinfo{author}{\bibfnamefont{P.}~\bibnamefont{Gartner}}, \bibnamefont{and}
  \bibinfo{author}{\bibfnamefont{F.}~\bibnamefont{Jahnke}},
  \bibinfo{journal}{Physica Status Solidi (b)} \textbf{\bibinfo{volume}{248}},
  \bibinfo{pages}{879} (\bibinfo{year}{2011}{\natexlab{a}}).

\bibitem[{\citenamefont{Gies et~al.}(2011{\natexlab{b}})\citenamefont{Gies,
  Florian, Gartner, and Jahnke}}]{Gies2011}
\bibinfo{author}{\bibfnamefont{C.}~\bibnamefont{Gies}},
  \bibinfo{author}{\bibfnamefont{M.}~\bibnamefont{Florian}},
  \bibinfo{author}{\bibfnamefont{P.}~\bibnamefont{Gartner}}, \bibnamefont{and}
  \bibinfo{author}{\bibfnamefont{F.}~\bibnamefont{Jahnke}},
  \bibinfo{journal}{Opt. Express} \textbf{\bibinfo{volume}{19}},
  \bibinfo{pages}{14370} (\bibinfo{year}{2011}{\natexlab{b}}).

\bibitem[{\citenamefont{Hennessy et~al.}(2007)\citenamefont{Hennessy, Badolato,
  Winger, Gerace, Atatüre, Gulde, Fält, Hu, and Imamoglu}}]{Imamoglu2007}
\bibinfo{author}{\bibfnamefont{K.}~\bibnamefont{Hennessy}},
  \bibinfo{author}{\bibfnamefont{A.}~\bibnamefont{Badolato}},
  \bibinfo{author}{\bibfnamefont{M.}~\bibnamefont{Winger}},
  \bibinfo{author}{\bibfnamefont{D.}~\bibnamefont{Gerace}},
  \bibinfo{author}{\bibfnamefont{M.}~\bibnamefont{Atatüre}},
  \bibinfo{author}{\bibfnamefont{S.}~\bibnamefont{Gulde}},
  \bibinfo{author}{\bibfnamefont{S.}~\bibnamefont{Fält}},
  \bibinfo{author}{\bibfnamefont{E.}~\bibnamefont{Hu}}, \bibnamefont{and}
  \bibinfo{author}{\bibfnamefont{A.}~\bibnamefont{Imamoglu}},
  \bibinfo{journal}{Nature} \textbf{\bibinfo{volume}{445}},
  \bibinfo{pages}{896} (\bibinfo{year}{2007}).

\bibitem[{\citenamefont{Kaniber et~al.}(2008)\citenamefont{Kaniber, Laucht,
  Neumann, Villas-B\^oas, Bichler, Amann, and Finley}}]{Finley2008}
\bibinfo{author}{\bibfnamefont{M.}~\bibnamefont{Kaniber}},
  \bibinfo{author}{\bibfnamefont{A.}~\bibnamefont{Laucht}},
  \bibinfo{author}{\bibfnamefont{A.}~\bibnamefont{Neumann}},
  \bibinfo{author}{\bibfnamefont{J.~M.} \bibnamefont{Villas-B\^oas}},
  \bibinfo{author}{\bibfnamefont{M.}~\bibnamefont{Bichler}},
  \bibinfo{author}{\bibfnamefont{M.-C.} \bibnamefont{Amann}}, \bibnamefont{and}
  \bibinfo{author}{\bibfnamefont{J.~J.} \bibnamefont{Finley}},
  \bibinfo{journal}{Phys. Rev. B} \textbf{\bibinfo{volume}{77}},
  \bibinfo{pages}{161303} (\bibinfo{year}{2008}).

\bibitem[{\citenamefont{Andr\'e et~al.}(2010)\citenamefont{Andr\'e, Jin,
  Brosco, Cole, Romito, Shnirman, and Sch\"on}}]{Cole2010}
\bibinfo{author}{\bibfnamefont{S.}~\bibnamefont{Andr\'e}},
  \bibinfo{author}{\bibfnamefont{P.-Q.} \bibnamefont{Jin}},
  \bibinfo{author}{\bibfnamefont{V.}~\bibnamefont{Brosco}},
  \bibinfo{author}{\bibfnamefont{J.~H.} \bibnamefont{Cole}},
  \bibinfo{author}{\bibfnamefont{A.}~\bibnamefont{Romito}},
  \bibinfo{author}{\bibfnamefont{A.}~\bibnamefont{Shnirman}}, \bibnamefont{and}
  \bibinfo{author}{\bibfnamefont{G.}~\bibnamefont{Sch\"on}},
  \bibinfo{journal}{Phys. Rev. A} \textbf{\bibinfo{volume}{82}},
  \bibinfo{pages}{053802} (\bibinfo{year}{2010}).

\bibitem[{\citenamefont{Carmichael}(1993)}]{carmichael1993}
\bibinfo{author}{\bibfnamefont{H.~J.} \bibnamefont{Carmichael}},
  \bibinfo{journal}{Phys. Rev. Lett.} \textbf{\bibinfo{volume}{70}},
  \bibinfo{pages}{2273} (\bibinfo{year}{1993}).

\bibitem[{\citenamefont{M{\o}lmer et~al.}(1993)\citenamefont{M{\o}lmer, Castin,
  and Dalibard}}]{Molmer1993}
\bibinfo{author}{\bibfnamefont{K.}~\bibnamefont{M{\o}lmer}},
  \bibinfo{author}{\bibfnamefont{Y.}~\bibnamefont{Castin}}, \bibnamefont{and}
  \bibinfo{author}{\bibfnamefont{J.}~\bibnamefont{Dalibard}},
  \bibinfo{journal}{J. Opt. Soc. Am. B} \textbf{\bibinfo{volume}{10}},
  \bibinfo{pages}{524} (\bibinfo{year}{1993}).

\bibitem[{\citenamefont{Johansson et~al.}(2012)\citenamefont{Johansson, Nation,
  and Nori}}]{Johansson2012}
\bibinfo{author}{\bibfnamefont{J.}~\bibnamefont{Johansson}},
  \bibinfo{author}{\bibfnamefont{P.}~\bibnamefont{Nation}}, \bibnamefont{and}
  \bibinfo{author}{\bibfnamefont{F.}~\bibnamefont{Nori}},
  \bibinfo{journal}{Computer Physics Communications}
  \textbf{\bibinfo{volume}{183}}, \bibinfo{pages}{1760 }
  (\bibinfo{year}{2012}).

\bibitem[{\citenamefont{Yamamoto and Imamo\ifmmode~\breve{g}\else
  \u{g}\fi{}lu}(1999)}]{yamamoto}
\bibinfo{author}{\bibfnamefont{Y.}~\bibnamefont{Yamamoto}} \bibnamefont{and}
  \bibinfo{author}{\bibfnamefont{A.}~\bibnamefont{Imamo\ifmmode~\breve{g}\else
  \u{g}\fi{}lu}}, \emph{\bibinfo{title}{Mesoscopic Quantum Optics}}
  (\bibinfo{publisher}{John Wiley \& Sons, Inc.}, \bibinfo{year}{1999}),
  \bibinfo{edition}{1st} ed.

\bibitem[{\citenamefont{M\o{}lmer}(1997)}]{moelmer1997}
\bibinfo{author}{\bibfnamefont{K.}~\bibnamefont{M\o{}lmer}},
  \bibinfo{journal}{Phys. Rev. A} \textbf{\bibinfo{volume}{55}},
  \bibinfo{pages}{3195} (\bibinfo{year}{1997}).

\bibitem[{\citenamefont{Loudon}(2000)}]{laudon}
\bibinfo{author}{\bibfnamefont{R.}~\bibnamefont{Loudon}},
  \emph{\bibinfo{title}{The Quantum Theory of Light}}
  (\bibinfo{publisher}{Oxford Science Publications}, \bibinfo{year}{2000}),
  \bibinfo{edition}{3rd} ed.

\bibitem[{\citenamefont{Elvira et~al.}(2011)\citenamefont{Elvira, Hachair,
  Verma, Braive, Beaudoin, Robert-Philip, Sagnes, Baek, Nam, Dauler
  et~al.}}]{elvira2011}
\bibinfo{author}{\bibfnamefont{D.}~\bibnamefont{Elvira}},
  \bibinfo{author}{\bibfnamefont{X.}~\bibnamefont{Hachair}},
  \bibinfo{author}{\bibfnamefont{V.~B.} \bibnamefont{Verma}},
  \bibinfo{author}{\bibfnamefont{R.}~\bibnamefont{Braive}},
  \bibinfo{author}{\bibfnamefont{G.}~\bibnamefont{Beaudoin}},
  \bibinfo{author}{\bibfnamefont{I.}~\bibnamefont{Robert-Philip}},
  \bibinfo{author}{\bibfnamefont{I.}~\bibnamefont{Sagnes}},
  \bibinfo{author}{\bibfnamefont{B.}~\bibnamefont{Baek}},
  \bibinfo{author}{\bibfnamefont{S.~W.} \bibnamefont{Nam}},
  \bibinfo{author}{\bibfnamefont{E.~A.} \bibnamefont{Dauler}},
  \bibnamefont{et~al.}, \bibinfo{journal}{Phys. Rev. A}
  \textbf{\bibinfo{volume}{84}}, \bibinfo{pages}{061802}
  (\bibinfo{year}{2011}).

\bibitem[{\citenamefont{Coldren and Corzine}(1995)}]{Coldren}
\bibinfo{author}{\bibfnamefont{L.~A.} \bibnamefont{Coldren}} \bibnamefont{and}
  \bibinfo{author}{\bibfnamefont{S.~W.} \bibnamefont{Corzine}},
  \emph{\bibinfo{title}{Diode Lasers and Photonic Integrated Circuits}}
  (\bibinfo{publisher}{Wiley-Interscience}, \bibinfo{year}{1995}).

\bibitem[{\citenamefont{Bjork and Yamamoto}(1991)}]{Yamamoto1991}
\bibinfo{author}{\bibfnamefont{G.}~\bibnamefont{Bjork}} \bibnamefont{and}
  \bibinfo{author}{\bibfnamefont{Y.}~\bibnamefont{Yamamoto}},
  \bibinfo{journal}{Quantum Electronics, IEEE Journal of}
  \textbf{\bibinfo{volume}{27}}, \bibinfo{pages}{2386 } (\bibinfo{year}{1991}).

\bibitem[{\citenamefont{Fricke}(1996)}]{fricke1996}
\bibinfo{author}{\bibfnamefont{J.}~\bibnamefont{Fricke}},
  \bibinfo{journal}{Annals of Physics} \textbf{\bibinfo{volume}{252}},
  \bibinfo{pages}{479 } (\bibinfo{year}{1996}).

\bibitem[{\citenamefont{Kira and Koch}(2006)}]{Kira2006}
\bibinfo{author}{\bibfnamefont{M.}~\bibnamefont{Kira}} \bibnamefont{and}
  \bibinfo{author}{\bibfnamefont{S.}~\bibnamefont{Koch}},
  \bibinfo{journal}{Progress in Quantum Electronics}
  \textbf{\bibinfo{volume}{30}}, \bibinfo{pages}{155 } (\bibinfo{year}{2006}).

\end{thebibliography}
\end{document}